\documentclass[journal]{IEEEtran} % 普通 trans 期刊模板

\usepackage{bm}
\usepackage{amsmath}
\usepackage{epsf}
\usepackage{graphics}
\usepackage{ amssymb }
\usepackage[dvips]{graphicx}
\usepackage{mathtools}% Loads amsmath
\usepackage{epsfig}
\usepackage{cite}
\usepackage[linesnumbered,ruled,vlined]{algorithm2e}
\usepackage{colortbl}
\usepackage{color}
\usepackage{hyperref} % 用于插入网页超链接
\usepackage{enumitem}
\usepackage{soul,xcolor}

\usepackage{bm}
\usepackage{amsmath}
\usepackage{epsf}
\usepackage{graphics}
\usepackage{ amssymb }
\usepackage[dvips]{graphicx}
\usepackage{epsfig}
\usepackage{cite}
\usepackage[linesnumbered,ruled,vlined]{algorithm2e}
\usepackage{graphicx}
\usepackage{epsfig}
\usepackage{latexsym}
\usepackage{amsfonts}
\usepackage{here}
\usepackage{rawfonts}
\usepackage[utf8]{inputenc}
\usepackage[english]{babel}
\usepackage{amsmath}
\usepackage{amsfonts}
\usepackage{amssymb}
\usepackage{color}
\usepackage{bm}
\usepackage{listings}
\usepackage{caption}
\usepackage{multicol}
\usepackage{tabularx,booktabs}
\newcolumntype{Y}{>{\centering\arraybackslash}X}
\makeatletter
\newcommand\notsotiny{\@setfontsize\notsotiny{6.31415}{7.1828}}
\makeatother
\usepackage{amssymb}
\usepackage{amsthm}
\usepackage{graphicx}
\usepackage{epstopdf}
\usepackage{listings}
\usepackage{float}
\usepackage{amsmath}
\usepackage{amssymb}
\usepackage{amsfonts}
\usepackage{epstopdf}

\usepackage{multirow}
\usepackage{amscd}
\usepackage{mathrsfs}
\usepackage{graphicx}
\usepackage{color}
\usepackage{url}
\usepackage{bm}
\usepackage{setspace}
\usepackage{footnote}
\usepackage{xcolor}
\definecolor{lightblue}{RGB}{173, 216, 230}  % 定义淡蓝色
\newtheorem{remark}{Remark}

\usepackage[]{algorithm2e}

\addto\captionsenglish{}
\usepackage{bbm}

\usepackage{multicol}
\usepackage{mathtools}

\usepackage{lipsum,graphicx,subcaption}

\usepackage{diagbox}
\usepackage{hyperref}
\usepackage[protrusion=true,expansion=true]{microtype}
\pdfoutput=1
\usepackage[font=footnotesize]{caption}
\allowdisplaybreaks[4]

\usepackage{graphicx}
\usepackage{grffile}
\usepackage{tabularx}
\newcounter{term}[section]

\renewcommand\theterm{\alph{term}}
\makeatletter
\newcommand{\vast}{\bBigg@{4}}

\newcommand{\Vast}{\bBigg@{5}}
\makeatother
\newcommand\semiHuge{\fontsize{22.7}{31.38}\selectfont}
\usepackage{soul,xcolor}

\usepackage[most]{tcolorbox}
\tcbset{colback=yellow!10!white, colframe=black!50!black, 
        highlight math style= {enhanced, %<-- needed for the ’remember’ options
            colframe=black,colback=red!10!white,boxsep=0pt}
        }
        \allowdisplaybreaks

\begin{document} 
\title{{\semiHuge A Lyapunov-Guided Diffusion-Based Reinforcement Learning Approach for UAV-Assisted Vehicular Networks with Delayed CSI Feedback}}
\author{Zhang Liu,~\IEEEmembership{Student Member,~IEEE}, Lianfen Huang,~\IEEEmembership{Member,~IEEE}, Zhibin Gao,~\IEEEmembership{Member,~IEEE}, \\ Xianbin Wang,~\IEEEmembership{Fellow,~IEEE}, Dusit Niyato,~\IEEEmembership{Fellow,~IEEE}, and Xuemin (Sherman) Shen,~\IEEEmembership{Fellow,~IEEE}
\thanks{\emph{Z. Liu (zhangliu@stu.xmu.edu.cn) is with the Department of Informatics and Communication Engineering, Xiamen University, Fujian, China 361102. L. Huang (lfhuang@xmu.edu.cn) is with the School of Information Science and Technology, Xiamen University Tan Kah Kee College, and with the Department of Informatics and Communication Engineering, Xiamen University, Fujian 361102, China. Z. Gao (gaozhibin@jmu.edu.cn) is with the Navigation Institute, Jimei University, Xiamen, Fujian, China 361021. X. Wang (xianbin.wang@uwo.ca) is with the Department of Electrical and Computer Engineering, Western University, London, Ontario N6A 5B9, Canada. D. Niyato (dniyato@ntu.edu.sg) is with the College of Computing and Data Science, Nanyang Technological University, Singapore. Xuemin (Sherman) Shen (sshen@uwaterloo.ca) is with the Department of Electrical and Computer Engineering, University of Waterloo, Waterloo, ON N2L 3G1, Canada. (Corresponding author: Lianfen Huang.)} }
} 
\maketitle
\vspace{-9mm}
\setulcolor{red}
\setul{red}{2pt}
\setstcolor{red}   

\begin{abstract}
Low altitude uncrewed aerial vehicles (UAVs) are expected to facilitate the development of aerial–ground integrated intelligent transportation systems and unlocking the potential of the emerging low-altitude economy. However, several critical challenges persist, including the dynamic optimization of network resources and UAV trajectories, limited UAV endurance, and imperfect channel state information (CSI). In this paper, we offer new insights into low-altitude economy networking by exploring intelligent UAV-assisted vehicle-to-everything communication strategies aligned with UAV energy efficiency. Particularly, we formulate an optimization problem of joint channel allocation, power control, and flight altitude adjustment in UAV-assisted vehicular networks. Taking CSI feedback delay into account, our objective is to maximize the vehicle-to-UAV communication sum rate while satisfying the UAV’s long-term energy constraint. To this end, we first leverage Lyapunov optimization to decompose the original long-term problem into a series of per-slot deterministic subproblems. We then propose a diffusion-based deep deterministic policy gradient (D3PG) algorithm, which innovatively integrates diffusion models to determine optimal channel allocation, power control, and flight altitude adjustment decisions. Through extensive simulations using real-world vehicle mobility traces, we demonstrate the superior performance of the proposed D3PG algorithm compared to existing benchmark solutions.

\end{abstract}
% \vspace{-.1mm}
\begin{IEEEkeywords}
Low-altitude economy networking, UAV-assisted vehicular networks, Lyapunov optimization, diffusion models, reinforcement learning, resource management, UAV trajectory planning.
\end{IEEEkeywords}
\vspace{-3mm}
\section{Introduction} \label{sec:intro}
\subsection{Background and Overview} \label{subsec:background}
\vspace{-.15mm}
With the rapid advancement of sensing and wireless technologies, vehicular networks have made significant strides, transforming traditional transportation systems into intelligent transportation systems (ITS)~\cite{khelifi2019named}. Nevertheless, effective ITS operation relies on dynamic vehicular communications with ubiquitous connectivity, low latency, and high reliability~\cite{cheng2020comprehensive}. By integrating various communication methods, such as vehicle-to-infrastructure (V2I) and vehicle-to-vehicle (V2V), vehicle-to-everything (V2X) technologies provide tailored support to ITS by meeting diverse quality of service (QoS) requirements of different vehicular communications~\cite{guo2020intelligent, he2023aerial}. Specifically, high-capacity V2I communications are engaged to deliver infotainment services (e.g., high-definition maps and augmented reality navigation), while high-reliability V2V communications ensure road safety services (e.g., cooperative driving and incident reporting).
 
However, guaranteeing seamless connectivity and providing uninterrupted services through integrated V2X communications present significant challenges. This difficulty is mainly due to the distinct technical challenges faced by V2I and V2V communications. On one hand, V2I infrastructures (e.g., base stations and roadside units) perform poorly in urban areas with high vehicle density~\cite{dai2023uav}, bringing excessive communication requests. As the number of vehicles increases, the V2I transmission rate decreases due to intense competition. Additionally, massive deployment of V2I infrastructures in rural areas and on cross-border highways is often economically unfeasible~\cite{samir2020leveraging, liu2023rfid}. On the other hand, due to high mobility and the Doppler effect, which lead to significant path loss and frequent blockages, V2V communications may experience degraded link quality and transmission interruptions~\cite{qi2022energy}. 

Recently, due to their advantages in flexible deployment, efficient transmission, and cost-effectiveness~\cite{li2023joint}, uncrewed aerial vehicles (UAVs) have become promising platforms for many emerging applications, driving the growth of the low-altitude economy~\cite{liu2025generative}. For instance, Amazon Prime Air uses UAVs to deliver packages to customers, addressing the challenge of last-mile delivery~\cite{mokhtari2022relaying}. More importantly, UAVs can serve as aerial base stations, making them an attractive complement to terrestrial infrastructure in V2X communications~\cite{niu2020space, liao2022energy}. Specifically, in a favorable aerial-terrestrial propagation environment, UAVs have a high likelihood of establishing line-of-sight (LoS) communication links with vehicles. Additionally, with their controllable mobility, UAVs can adjust their positions to follow moving vehicles that need to establish connections.

\vspace{-3mm}
\subsection{Motivation and Main Challenges} \label{subsec:challenges}
\vspace{-.15mm}

Despite the above advantages, UAV-assisted vehicular networks still face several critical challenges that need to be carefully addressed. \emph{First, efficiently managing network resources and planning UAV trajectories under real-time dynamics can be very challenging.} UAV-assisted vehicular networks face spectrum scarcity due to the exponentially increasing number of different communication links~\cite{qi2022energy}. Co-channel interference between vehicle-to-UAV (V2U) and V2V links, caused by dynamic spectrum sharing, must be carefully managed to meet the desired QoS requirements. Specifically, the deeply coupled channel allocation and transmit power control dramatically increase the complexity of joint optimization under highly dynamic vehicular communications. Furthermore, determining UAV trajectories in spatiotemporal domains adds complexity to the issue. While the UAV can move closer to target vehicles to increase the probability of a LoS connection, it may become too distant from others, resulting in unstable connections with lower data rates.

\emph{Second, balancing the trade-off between the different network operational intents or objectives, including system communication performance and the energy consumption of UAVs, is challenging.} Since UAVs have limited endurance due to their finite battery life, it is crucial to ensure sustained operation in order to improve system performance~\cite{dai2023uav}. In other words, if a UAV consumes too much energy (e.g., by frequently adjusting its altitude to establish LoS connections), the available energy may be insufficient for subsequent service provisioning. As a result, the long-term performance would be significantly degraded. However, considering the dynamic characteristic of UAV-assisted vehicular networks (e.g., vehicle mobility and time-varying channel conditions), making decisions on optimization variables across consecutive time slots without knowledge of future dynamics is a non-trivial challenge.

\emph{Third, dynamic network operation with low-latency knowledge of channel state information (CSI) from all communication links is challenging.} When UAVs serve as aerial base stations, they can only estimate the CSI between vehicles and UAVs~\cite{he2023aerial}. In this case, the CSI of V2V links is reported to the aerial base stations periodically, potentially causing additional CSI feedback delays. Given the rapidly time-varying channel characteristics in high-speed vehicular networks (e.g., Doppler shift and multipath fading), channel allocation, power control, and UAV trajectory planning schemes based on periodically reported CSI will inevitably degrade system performance. This happens due to severe mismatches between the actual channel state and the estimated CSI, leading to reduced throughput and a higher probability of link interruptions.

\emph{Fourth, conventional optimization methods are unsuitable for dynamic vehicular networks.} In practical scenarios, vehicle locations and wireless channel conditions change over time, meaning an optimal solution derived for one specific moment and situation may not remain optimal in the long run. Furthermore, convex optimization methods~\cite{he2023aerial} and heuristic algorithms~\cite{mokhtari2022relaying} either rely on problem convexity, leading to extensive computations, or require numerous iterations to converge, making it challenging to meet real-time decision-making requirements. Recently, deep reinforcement learning (DRL)~\cite{zhang2024performance, liu2024ga} has emerged as a promising technique for real-time decision-making by learning the relationship between input states (e.g., vehicle mobility) and actions (e.g., channel allocation). Unfortunately, DRL also faces challenges in balancing exploration and exploitation—excessive exploration may lead to suboptimal solutions, while excessive exploitation can result in short-sighted decision-making.

\vspace{-3mm}
\subsection{Summary of Contributions} \label{subsec:contributions}
\vspace{-.15mm}

Motivated by the above challenges, we formulate a joint optimization problem involving channel allocation, power control, and trajectory planning for UAV-assisted vehicular networks. The objective is to maximize the V2U communication sum rate while ensuring the UAV’s long-term energy constraint. Our main contributions are as follows:

\begin{itemize}[leftmargin=4mm]
\item \textbf{\textit{Framework:}} We consider the CSI feedback delay and formulate the joint channel allocation, power control, and flight altitude adjustment problem in UAV-assisted vehicular networks as a mixed-integer nonlinear programming (MINLP) problem, which is known to be NP-hard. This makes the problem particularly challenging to solve, especially in the presence of vehicle mobility, time-varying channel conditions, and the UAV's long-term energy constraint.
\item \textbf{\textit{Solution:}} To this end, we first employ the Lyapunov optimization technique to decouple the original problem into a series of per-slot deterministic subproblems, ensuring the UAV's sustained operation under stochastic conditions. Building on this transformation, we propose a diffusion-based deep deterministic policy gradient (D3PG) algorithm to address the problem on a per-slot basis.
\item \textbf{\textit{Innovation:}} In the D3PG algorithm, we leverage diffusion models—originally developed for image generation—to optimize channel allocation, power control, and UAV flight altitude adjustment decisions. The denoising process in diffusion models effectively addresses the exploration–exploitation trade-off in DRL and enhances decision robustness in dynamic environments with CSI feedback delay.
\item \textbf{\textit{Validation:}} We design our simulation scenario based on a real-world road network extracted from OpenStreetMap~\cite{haklay2008openstreetmap} and use SUMO~\cite{lopez2018microscopic} to simulate vehicle mobility, thereby establishing a realistic UAV-assisted vehicular networks. We then evaluate the effectiveness of the proposed D3PG algorithm through experiments under various simulation settings, comparing its performance with three benchmark solutions.

% , demonstrating that our method not only achieves a higher communication rate but also satisfies the long-term energy constraint for sustainable UAV operation.
\end{itemize}

\vspace{-3mm}
\subsection{Paper Organization} \label{subsec:organization}
\vspace{-.15mm}
The rest of the paper is structured as follows: Sec.~\ref{sec:related_works} reviews related works. Sec.~\ref{sec:systemmodel_problemformulation} describes the system model and formulates the joint optimization problem of channel allocation, power control, and flight altitude adjustment in UAV-assisted vehicular networks. Sec.~\ref{sec:Lyapunov} proposes Lyapunov optimization technique to handle the original problem. Sec.~\ref{sec:diffusion_model} introduces the preliminaries of the diffusion model. Sec.~\ref{sec:D3PG} presents our proposed D3PG algorithm. Sec.~\ref{sec:simulation} details the simulation results, followed by the conclusion and future work in Sec.~\ref{sec:conclusion}.

\vspace{-3mm}
\section{Related Work} \label{sec:related_works}
\vspace{-.15mm}
Henceforth, we summarize the contributions of related works and highlight the aspects they have not addressed, which serve as the primary motivations for this work.

\vspace{-3mm}
\subsection{UAV-Assisted Communications for Static Ground Users}\label{subsec:UAV_ground_users}
\vspace{-.15mm}

UAVs have been extensively studied and utilized in the literature as flying base stations and relay nodes to enhance communication quality for ground users. The authors in~\cite{10138117} investigated the joint optimization of UAV trajectory and resource allocation, aiming to maximize system energy efficiency while ensuring the service quality of all ground users. The authors in~\cite{9800925} explored the joint optimization of the number and placement of UAVs to ensure wireless coverage for all ground users. The authors in~\cite{9714909} studied uplink transmission in a UAV-assisted cellular network, aiming to minimize the transmit power consumption of both users and UAVs. The authors in~\cite{9454157} proposed an online data-driven multi-UAV trajectory and transmission control scheme to optimize the quality-of-experience for ground users.

Although these works achieve satisfactory performance in their respective scenarios, they assume a deterministic and static user distribution. When considering highly dynamic vehicular networks with time-varying channel conditions and stochastic vehicle movements, the aforementioned schemes face various technical challenges in performance optimization, necessitating further exploration.

\vspace{-3mm}
\subsection{UAV-Assisted Vehicular Networks}\label{subsec:UAV_vehicular_networks}
\vspace{-.15mm}

As a highly mobile and easily deployable facility, the UAV is well-suited for communications in dynamic vehicular networks. The authors in~\cite{mokhtari2022relaying} addressed the power and data rate allocation problem in UAV-enabled vehicular ad-hoc networks, aiming to minimize communication delay while maximizing energy efficiency. The authors in~\cite{zhang2024performance} studied the UAV's 3D position deployment problem to analyze system performance in terms of the vehicular users' successful service probability. The authors in~\cite{li2023joint} introduced a UAV-aided relaying system for vehicular networks, aiming to reduce transmission time by jointly optimizing relay selection and transmission scheduling. The authors in~\cite{zhang2021uav} proposed a novel UAV-enabled scheduling protocol for vehicular networks to enhance the efficiency of V2X data dissemination.

However, these works either implicitly assume perfect CSI acquisition~\cite{mokhtari2022relaying,zhang2024performance}, overlooking the CSI feedback delay in dynamic vehicular networks caused by Doppler shift and multipath fading, or neglect the UAV’s long-term energy constraint due to its finite battery life~\cite{li2023joint,zhang2021uav}, which can significantly impact its sustained operation for long-term service provisioning.

\vspace{-3mm}
\subsection{Usage of Deep Reinforcement Learning in Optimization}\label{subsec:DRL_optimization}
\vspace{-.15mm}

Recently, learning-based algorithms, particularly DRL, have been widely applied to improve real-time decision-making and solution design for complex optimization problems. The authors in~\cite{9916069} proposed a DRL-based UAV path planning scheme that learns the historical locations of different cluster heads to determine optimal hover points for the UAV. The authors in~\cite{9748970} employed a double deep Q-network with a dueling architecture to assist the UAV in determining the optimal flying direction for each time slot. The authors in~\cite{9465671} introduced a deep Q-network framework combined with a difference-of-convex algorithm to jointly optimize UAV positioning and radio resource allocation. The authors in~\cite{10195210} explored a multi-agent two-timescale DRL algorithm for power allocation and content placement of content providers, aiming to enhance delivery success probability and content hit ratio.
 
Although DRL utilizes deep neural networks (DNNs) to learn the relationship between a problem's state space (e.g., vehicle mobility) and its action space (e.g., channel allocation), making it well-suited for real-time decision-making in dynamic vehicular networks, the use of common multi-layer perceptron (a type of fully connected DNN) in DRL architectures is ineffective due to the exploration-exploitation trade-off and the risk of converging to suboptimal policies~\cite{10736570}.

\vspace{-3mm}
\section{System Model and Problem Formulation} \label{sec:systemmodel_problemformulation}
\vspace{-.15mm}

In this section, we first provide an overview of the network, detailing the UAV-assisted vehicular networks considered in this paper. We then introduce the V2U and V2V collaborative communication models, followed by the UAV energy consumption model. Consequently, we formulate the joint optimization problem of channel allocation, power control, and flight altitude adjustment to maximize the V2U communication sum rate while ensuring the UAV’s long-term energy constraint.

\vspace{-3mm}
\subsection{Network Outline}\label{subsec:system_architecture}
\vspace{-.15mm}

\begin{figure}[t!]
\includegraphics[width=.48\textwidth]{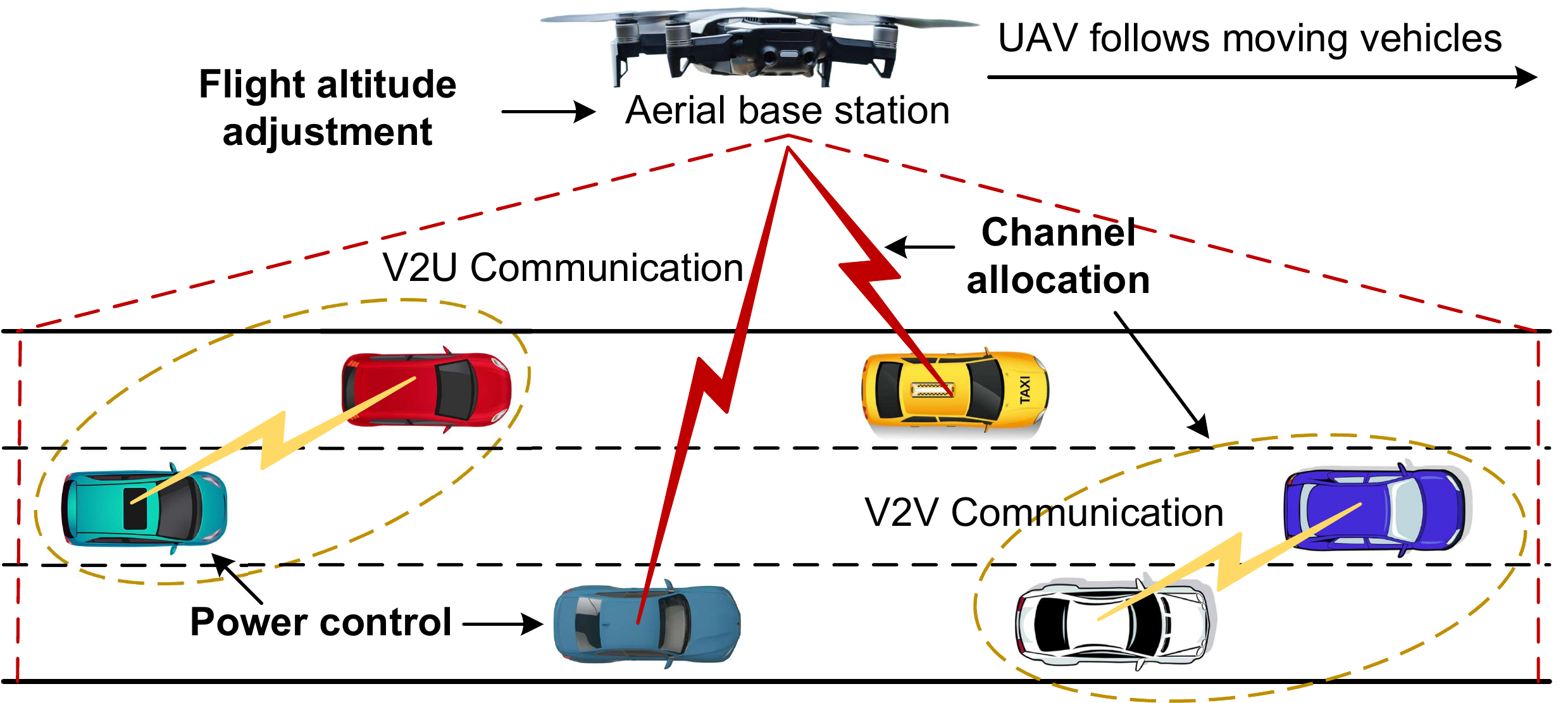}
\centering
\vspace{-1.5mm}
\caption{A schematic illustration of UAV-assisted vehicular networks incorporating both V2U and V2V communication links.}
\label{fig:uav_vehicular_networks}
% \vspace{-1mm}
\end{figure}

Fig.~\ref{fig:uav_vehicular_networks} illustrates the UAV-assisted vehicular network of interest, consisting of a single UAV acting as an aerial base station and several moving vehicles. Specifically, we consider a unidirectional highway scenario that lacks terrestrial infrastructure due to remoteness or post-disaster conditions. The network system operates over a time window divided into discrete time slots, denoted as $\mathcal{T}=\{1, \dots,T\}$. A standalone UAV moves at a constant speed, following the vehicles to provide communication services.\footnote{In this work, we consider a single UAV for simplicity. However, our approach can be extended to a multi-UAV scenario by dividing the highway into several segments, each serviced by a separate UAV.} Leveraging cellular technology, vehicles can upload their sensing data to the UAV to enable collaborative sensing services via V2U communications, where the set of V2U communication links is denoted as $\mathcal{M}=\{1, \dots,M\}$. Additionally, leveraging device-to-device communication technology, vehicles can establish V2V connections to exchange real-time local data for incident reporting. The set of V2V communication links is denoted as $\mathcal{K}=\{1, \dots,K\}$, where $K \leq M$.

In this work, we adopt orthogonal frequency division multiplexing (OFDM) modulation, dividing the spectrum into $M$ orthogonal channels, where $M$ V2U communication links are pre-allocated to operate separately over these channels~\cite{he2023aerial, tian2021multiagent}. To enhance spectrum utilization efficiency, the orthogonal channels allocated for V2U communications can be shared with V2V pairs. While spectrum sharing increases network flexibility and scalability, proper resource orchestration is essential to mitigate co-channel interference. To this end, we introduce a binary variable $x_{k,m}(t)$ to represent the \emph{channel allocation} policy for V2V communications at time slot $t$, where $x_{k,m}(t)=1$ indicates that the $k$-th V2V link shares the same spectrum with the $m$-th V2U link at time slot $t$; otherwise $x_{k,m}(t)=0$. Note that each V2V pair can occupy only a single channel for data transmission in any given time slot. For ease of reference, key notations used in the article are summarized in Table~\ref{table1}.

\begin{table}[!t]
\vspace{-2mm}
\centering
\footnotesize
\caption{Summary of Key Notations.}
\rowcolors{1}{white}{gray!20}
\label{table1}
% \rowcolors{1}{white}{gray!20}
% \rowcolors{1}{lightblue!30}{white}
% \scalebox{0.8}{\small
\begin{tabular}{|c|p{6.3cm}|}
\hline
% \rowcolor{lightgray} % 设置这一行的背景色为浅灰色
\textbf{Notations}                    & \textbf{Description}   \\ \hline \hline
$\widehat{g^{\text{V}}_{k}}(t)$  & Small-scale fading between V2V communication pair $k$ at time slot $t$, prior to the feedback delay             \\ \hline
$\widehat{g^{\text{V}}_{m,k}}(t)$                  & Small-scale fading from V2U transmitter $m$ to V2V receiver $k$ at time slot $t$, prior to the feedback delay          \\ \hline
% $\Delta H(t)$                          & UAV flight altitude adjustment decision at time slot $t$        \\\hline
$h^{\text{U}}_{m}(t)$   & Uplink channel gain from V2U transmitter $m$ to the UAV at time slot $t$          \\ \hline
$h^{\text{U}}_{k}(t)$                 & Uplink channel gain from V2V transmitter $k$ to the UAV at time slot $t$   \\ \hline
$h^{\text{V}}_{k}(t)$                  & Uplink channel gain between V2V communication pair $k$ at time slot $t$                \\ \hline
$h^{\text{V}}_{m,k}(t)$                  & Channel gain from V2U transmitter $m$ to V2V receiver $k$ at time slot $t$      \\ \hline
$\mathcal{K}$                    & Index set of V2V communication links               \\ \hline
$\mathcal{M}$                  & Index set of V2U communication links                \\ \hline
$P(t)$                  & UAV flight power consumption at time slot $t$     \\ \hline
% $p_{m}(t),p_k(t)$                          & Power control decision for V2U communication $m$, and V2V communication $k$ at time slot $t$, respectively         \\\hline 
$Q(t)$              &  Virtual queue for UAV flight energy consumption at time slot $t$         \\ \hline
$\mathcal{T}$          & Index set of time slots       \\ \hline
% $R^{\text{U}}_{m}(t)$                  & V2U communication sum rate at time slot $t$         \\ \hline
% \textbf{Notations in Optimization Parameters}                    & \textbf{Description}   \\ \hline \hline
% $x_{k,m}(t)$                  & Channel allocation decision for V2V communication link $k$ at time slot $t$         \\ \hline
\end{tabular}
% }
% \vspace{-5mm}
\end{table}

\vspace{-3mm}
\subsection{V2U Communication Model}\label{subsec:v2u_communication}
\vspace{-.15mm}
To evaluate the uplink performance of V2U communications, we model the signal-to-interference-plus-noise ratio (SINR) of the $m$-th V2U link at time slot $t$ as
\vspace{-.5mm}
\begin{equation}\label{eq:v2u_sinr}
\gamma^{\text{U}}_{m}(t)= \frac{p_m(t)h^{\text{U}}_{m}(t)}{\sum^{K}_{k=1} \Big(x_{k,m}(t)p_k(t)h^{\text{U}}_{k}(t) \Big)+N_0B},
\end{equation}
where $p_m(t)$ and $p_k(t)$ denote the \emph{transmit powers} of V2U transmitter $m$ and V2V transmitter $k$, respectively,\footnote{For simplicity of expression, we refer to the transmitting vehicles for V2U communication $m$ and V2V communication $k$ as V2U transmitter $m$ and V2V transmitter $k$, respectively, hereafter.} $N_0$ is the noise power spectral density, and $B$ is the bandwidth of each channel. Additionally, the uplink channel gain $h^{\text{U}}_{m}(t)$ from V2U transmitter $m$ to the UAV at time slot $t$ is given by
\vspace{-.5mm}
\begin{equation}\label{eq:v2u_gain_v2u}
h^{\text{U}}_{m}(t)=\frac{|g^{\text{U}}_{m}(t)|^2}{\text{PL}^{\text{U}}_{m}(t)},
\end{equation}
where $g^{\text{U}}_{m}(t)\sim \mathcal{CN}(0,1)$ represents the small-scale fading, and $\text{PL}^{\text{U}}_{m}(t)$ denotes the large-scale path loss from V2U transmitter $m$ to the UAV at time slot $t$.

Then, the path loss $\text{PL}^{\text{U}}_{m}(t)$ considers both line-of-sight (LoS) and non-line-of-sight (NLoS) components. Specifically, it is expressed as a weighted sum: 
\vspace{-.5mm}
\begin{equation}\label{eq:average_path_loss}
\text{PL}^{\text{U}}_{m}(t)=\text{Pr}_{\text{LoS}}\text{PL}^{\text{U,LoS}}_{m}(t)+(1-\text{Pr}_{\text{LoS}})\text{PL}^{\text{U,NLoS}}_{m}(t),
\end{equation}
where $\text{Pr}_{\text{LoS}}$ is the probability of a LoS connection, and $\text{PL}^{\text{U,LoS}}_{m}(t)$ and $\text{PL}^{\text{U,NLoS}}_{m}(t)$ represent the path losses under LoS and NLoS conditions (expressed in dB), respectively, which can be given by
\vspace{-.5mm}
\begin{align}\label{eq:path_loss}
\text{PL}^{\text{U,LoS}}_{m}(t)&=20\log_{10}\frac{4\pi f_cd^{\text{U}}_m(t)}{c} +\alpha_{\text{LoS}}, \\
\text{PL}^{\text{U,NLoS}}_{m}(t)&=20\log_{10}\frac{4\pi f_cd^{\text{U}}_m(t)}{c} +\alpha_{\text{NLoS}},
\end{align}
where $f_c$ is the carrier frequency, $c$ is the speed of light, and $\alpha_{\text{LoS}}$, $\alpha_{\text{NLoS}}$ are the mean additional losses under LoS and NLoS conditions, respectively. Moreover, $d_m(t)=\sqrt{H(t)^2 + |\bm{l}_{\text{U}}(t) - \bm{l}_m(t)|^2}$ is the 3D distance between the UAV and V2U transmitter $m$, where $H(t)$ denotes the \emph{flight altitude} of the UAV at time slot $t$, and $\bm{l}_{\text{U}}(t)$, $\bm{l}_m(t)$ represent the horizontal locations of the UAV and V2U transmitter $m$ at time slot $t$, respectively.  

Besides, the LoS probability is modeled as a function of the elevation angle~\cite{yan2019comprehensive}:
\vspace{-.5mm}
\begin{equation}\label{eq:los_prob}
\text{Pr}_{\text{LoS}}=\frac{1}{1+a\text{exp}\Big(-b[\frac{180}{\pi}\theta-a]\Big)},
\end{equation}
where $a$ and $b$ are environment-dependent constants, and $\theta=\text{tan}^{-1}(\frac{H(t)}{||\bm{l}_{\text{U}}(t)-\bm{l}_m(t)||})$ is the elevation angle. Similarly, the channel gain $h^{\text{U}}_{k}(t)$ from V2V transmitter $k$ to the UAV at time slot $t$ can be modeled in the same way.

Finally, based on the SINR in~\eqref{eq:v2u_sinr}, the uplink data rate of the $m$-th V2U link is calculated via the Shannon formula as:
% \vspace{-.5mm}
\begin{equation}\label{eq:v2u_rate}
R^{\text{U}}_m(t) = B\text{log}_2\Big(1+\gamma^{\text{U}}_{m}(t)\Big).
\end{equation}

\vspace{-3mm}
\subsection{V2V Communication Model}\label{subsec:v2v_communication}
\vspace{-.15mm}

The SINR of the $k$-th V2V link at time slot $t$, denoted as $\gamma^{\text{V}}_{k}(t)$, is given by 
\vspace{-.5mm}
\begin{equation}\label{eq:v2v_sinr}
\hspace{-1mm} \gamma^{\text{V}}_{k}(t)= \frac{p_k(t)h^{\text{V}}_{k}(t)}{ \sum^{M}_{m=1} \Big(x_{k,m}(t)p_m(t)h^{\text{V}}_{m,k}(t)\Big)+N_0B},
\end{equation}
where the uplink channel gain $h^{\text{V}}_{k}(t)$ between V2V communication pair $k$ at time slot $t$ is also modeled by combining large-scale path loss and Rayleigh small-scale fading as
\vspace{-.5mm}
\begin{equation}\label{eq:v2u_gain_v2v}
h^{\text{V}}_{k}(t)=\frac{|g^{\text{V}}_{k}(t)|^2}{\text{PL}^{\text{V}}_{k}(t)}.
\end{equation}
Here, $g^{\text{V}}_{k}(t)\sim \mathcal{CN}(0,1)$ represents the small-scale fading, and $\text{PL}^{\text{V}}_{k}(t)=44.23+16.7\log_{10}||\bm{l}^{\text{Tx}}_k(t) - \bm{l}^{\text{Rx}}_k(t)||$~\cite{zhang2019deep} denotes the large-scale path-loss (expressed in dB) between V2V communication pair $k$ at time slot $t$, where $\bm{l}^{\text{Tx}}_k(t)$ and $\bm{l}^{\text{Rx}}_k(t)$ represent the horizontal locations of the V2V transmitter and receiver $k$, respectively. Similarly, the channel gain $h^{\text{V}}_{m,k}(t)$ from V2U transmitter $m$ to V2V receiver $k$ at time slot $t$ can be modeled in the same way.

% Moreover, term (I) in~\eqref{eq:v2v_sinr} indicates the interference from other V2V links ($k^{\prime} \in \mathcal{K}\setminus \{k\}$) sharing the same channel as the $k$-th V2V link at time slot $t$, where $h^{\text{V}}_{k^{\prime},k}(t)$ and $\alpha^{\text{V}}_{k^{\prime},k}(t)$ represent the Rayleigh small-scale fading and large-scale path-loss fading between the transmitting vehicle of the $k^{\prime}$-th V2V link and the receiving vehicle of the $k$-th V2V link at time slot $t$, respectively.

% \begin{table*}[!t]
% \begin{equation}\label{eq:v2v_sinr}
% \gamma^{\text{V}}_{k}(t)= \frac{p_k(t)|h^{\text{V}}_{k}(t)|^2 \alpha^{\text{V}}_{k}(t)}{ \sum^{M}_{m=1} \Big(x_{k,m}(t)p_m(t)|h^{\text{V}}_{m}(t)|^2 \alpha^{\text{V}}_{m}(t)\Big)+\underbrace{\sum_{k^{\prime} \in \mathcal{K}\setminus \{k\}}\sum^{M}_{m=1}x_{k^{\prime},m}(t)x_{k,m}(t)  \Big(p_{k^{\prime}}(t)|h^{\text{V}}_{k^{\prime},k}(t)|^2 \alpha^{\text{V}}_{k^{\prime},k}(t)\Big)}_{(\text{I})}+N_0B}
% \end{equation}
% \hrule
% \end{table*}

\begin{figure}[t!]
\includegraphics[width=.45\textwidth]{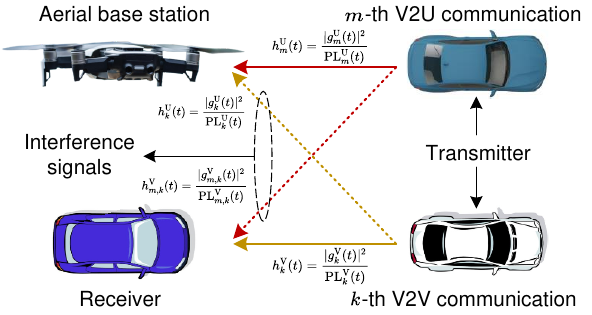}
\centering
\vspace{-1.5mm}
\caption{A schematic illustration of the V2U and V2V channel gains, along with their respective interference signals, in UAV-assisted V2X communications.}
\label{fig:csi_feedback_delay}
% \vspace{-1mm}
\end{figure}

However, given the rapidly time-varying channel characteristics in high-speed vehicular networks, obtaining accurate CSI is challenging. Specifically, as shown in Fig.~\ref{fig:csi_feedback_delay}, aside from $h^{\text{U}}_{m}(t)$ and $h^{\text{U}}_{k}(t)$, which can be directly obtained by the UAV (i.e., aerial base station), the CSI of V2V communications--namely, $h^{\text{V}}_{m,k}(t)$ and $h^{\text{V}}_{k}(t)$--is periodically reported to the UAV, requiring CSI estimation that accounts for additional feedback delays. Subsequently, we model the channel variation over a feedback delay $T_{\text{delay}}$, using the first order Gauss-Markov process, which can be given by~\cite{liang2017spectrum}
\vspace{-.5mm}
\begin{equation}\label{eq:channel_variation}
g(t)=J_0(2\pi \frac{f_cs_{\text{rel}}}{c}T_{\text{delay}})\hat{g}(t)+\delta,
\end{equation}
where $J_0(\cdot)$ denotes the zero-order Bessel function of the first kind, $g(t)$ represents the estimated small-scale fading in the current time slot $t$,\footnote{In this paper, we consider only the impact of CSI feedback delay on small-scale fading, as large-scale path loss changes gradually and remains relatively stable over short time intervals.} and $\hat{g}(t) \sim \mathcal{CN}(0,1)$ corresponds to the small-scale fading prior to the feedback delay. Additionally, $s_{\text{rel}}$ represents the relative vehicle speed and $\delta \sim \mathcal{CN} \Big(0,1-\Big[J_0(2\pi \frac{f_cs_{\text{rel}}}{c}T_{\text{delay}})\Big]^2 \Big)$ is the distribution of the channel discrepancy term.

Finally, based on~\eqref{eq:channel_variation}, $g^{\text{V}}_{m,k}(t)$ and $g^{\text{V}}_{k}(t)$ can be rewritten as 
% \vspace{-.5mm}
% \begin{align}\label{eq:h_kk}
% |h^{\text{V}}_{k^{\prime},k}(t)|^2 = \Big[J_0(2\pi \frac{f_cs_{\text{rel}}}{c}T_{\text{delay}}) \Big]^2|\widehat{h^{\text{V}}_{k^{\prime},k}}(t)|^2+(\delta^{\text{V}}_{k^{\prime},k})^2,
% \end{align}
\vspace{-.5mm}
\begin{align}\label{eq:h_m}
\!\!\! |g^{\text{V}}_{m,k}(t)|^2 &= \Big[J_0(2\pi \frac{f_cs_{\text{rel}}}{c}T_{\text{delay}}) \Big]^2|\widehat{g^{\text{V}}_{m,k}}(t)|^2+(\delta^{\text{V}}_{m,k})^2,\\
|g^{\text{V}}_{k}(t)|^2 &= \Big[J_0(2\pi \frac{f_cs_{\text{rel}}}{c}T_{\text{delay}}) \Big]^2|\widehat{g^{\text{V}}_{k}}(t)|^2+(\delta^{\text{V}}_{k})^2.
\end{align}

\vspace{-3mm}
\subsection{UAV Energy Consumption Model}\label{subsec:uav_energy_consumption}
\vspace{-.15mm}

The UAV's power consumption is crucial in UAV-assisted vehicular networks due to its limited battery capacity. In this paper, considering that the communication power of the UAV is negligible compared to its flight power~\cite{al2020uav}, we focus solely on the UAV's flight power consumption for simplicity, which can be expressed as~\cite{cai2022resource}
\vspace{-.5mm}
\begin{align}\label{eq:uav_energy_consumption}
\hspace{-1mm} P(t)&=\underbrace{P_0\Big(1+\frac{3\Big(v_x(t)^2+v_y(t)^2\Big)}{\Omega^2r^2} \Big)}_{\text{Blade profile power}}+\underbrace{\frac{P_1v_0}{v_x(t)^2+v_y(t)^2}}_{\text{Induced power}} \nonumber\\
&+\underbrace{\frac{1}{2}d_0\rho s_{r}A_r\Big(v_x(t)^2+v_y(t)^2 \Big)^{\frac{3}{2}}}_{\text{Parasite power}}+\underbrace{Gv_z(t)}_{\text{Vertical flight power}}\hspace{-3mm},
\end{align}
where the UAV's velocity in the 3D Cartesian coordinate system is represented as $\Big[v_x(t), v_y(t), v_z(t)\Big]^\top \in \mathbb{R}^{3 \times 1}$; $P_o$ denotes the blade profile power during hovering; $\Omega$ is the blade's angular velocity; $r$ is the rotor radius; $P_1$ is the induced power during hovering; $v_0$ is the induced velocity of the rotor during forward flight; $d_0$ is the fuselage drag ratio; $\rho$ is the air density; $s_r$ is the rotor solidity; $A_r$ is the rotor disc area, and $G$ represents the UAV's weight.

\vspace{-3mm}
\subsection{Problem Formulation}\label{subsec:prob_formulate}
\vspace{-.15mm}

We now formulate the joint channel allocation, power control, and flight altitude adjustment problem in UAV-assisted vehicular networks as a dynamic long-term optimization. Our objective is to maximize the V2U communication sum rate in~\eqref{eq:v2u_rate} for all V2U communication links across all time slots. This problem is formally defined as $\textbf{P1}$ below:
\vspace{-1.5mm}
\begin{align}
     &\textbf{P1}: \quad\quad\quad \max_{\{\bm{x},\bm{p}, \bm{\Delta H}\}} \frac{1}{TM}\sum_{t\in\mathcal{T}}\sum_{m\in\mathcal{M}} R^{\text{U}}_{m}(t)\hspace{-40mm}\label{eq:problem1} \\
     \text{s.t.} \quad
     &\mathcal{C}1: \  x_{k,m}(t) \in \{0,1\}, \ \forall k\in \mathcal{K}, m\in \mathcal{M}, t\in \mathcal{T},  \nonumber \\
     &\mathcal{C}2: \ p_m(t) \in [0,p_{\text{max}}], \ \forall m\in \mathcal{M}, t\in \mathcal{T},  \nonumber\\
     &\mathcal{C}3:  \ p_k(t) \in [0,p_{\text{max}}], \ \forall k\in \mathcal{K}, t\in \mathcal{T}, \nonumber\\ 
     &\mathcal{C}4:  \ H(t) \in [H_{\text{min}},H_{\text{max}}],  \ \forall t\in \mathcal{T}, \nonumber\\
     &\mathcal{C}5:  \ \sum^M_{m = 1}x_{k,m}(t)\leq 1,  \ \forall k\in \mathcal{K}, t\in \mathcal{T},  \nonumber\\
     &\mathcal{C}6:  \ \sum^K_{k = 1}x_{k,m}(t)\leq 1,  \ \forall m\in \mathcal{M}, t\in \mathcal{T},  \nonumber\\
     &\mathcal{C}7:  \ \text{Pr}\Big\{\gamma^{\text{V}}_k(t) < \gamma^{\text{V}}_{\text{th}} \Big\} \leq \text{Pr}^{\text{V}}_{\text{th}},  \ \forall k\in \mathcal{K}, t\in \mathcal{T},  \nonumber\\
     &\mathcal{C}8:  \ \lim_{T \to \infty} \frac{1}{T}\sum^T_{t=1}\mathbb{E}\{P(t) \Delta \} \leq E^{\text{U}}_{\text{th}}, \nonumber 
\end{align}
where $\bm{x}=\{x_{k,m}(t)\}_{k\in \mathcal{K},m\in\mathcal{M},t\in \mathcal{T}}$ represents the channel allocation vector for V2V communications reusing the spectrum of V2U communications, $\bm{p}=\{p_{m}(t),p_k(t)\}_{k\in \mathcal{K},m\in\mathcal{M},t\in \mathcal{T}}$ denotes the power control vector for the transmitting vehicles of both V2U and V2V communications, and $\bm{\Delta H}=\{\Delta H(t)\}_{t\in \mathcal{T}}$ denotes the UAV's flight altitude adjustment vector.

In $\textbf{P1}$, constraint $\mathcal{C}1$ ensures that the channel allocation decision is binary. Constraints $\mathcal{C}2$ and $\mathcal{C}3$ limit the maximum transmission power of V2U and V2V communications, respectively, with $p_{\text{max}}$ denoting the maximum vehicular transmission power. Constraint $\mathcal{C}4$ defines the value range for the UAV's flight altitude, where $H_{\text{min}}$ and $H_{\text{max}}$ represent the minimum and maximum UAV height limits, respectively. Constraint $\mathcal{C}5$ enforces exclusive spectrum access, permitting each V2V pair to utilize only one V2U link's spectrum. Complementarily, constraint $\mathcal{C}6$ ensures that each V2U link's spectrum can be shared with at most one V2V pair. Constraint $\mathcal{C}7$ guarantees the reliability of V2V communications, where $\gamma^{\text{V}}_{\text{th}}$ represents the minimum SINR required for V2V communications, and $\text{Pr}^{\text{V}}_{\text{th}}$ is the tolerated outage probability. Finally, constraint $\mathcal{C}8$ enforces a long-term UAV propulsion energy constraint to ensure the UAV's operational endurance, where $\Delta$ represents the duration of each time slot, and $E^{\text{U}}_{\text{th}}$ denotes the maximum allowed operational power of the UAV.

\vspace{-.5mm}
\begin{remark}\label{rem:NPhard} Due to the non-linearity and recursive nature of the long-term constraint $\mathcal{C}8$, the channel allocation, power control, and UAV flight altitude adjustment decisions are interdependent over time. Additionally, due to the presence of both discrete and continuous variables as defined by constraints $\mathcal{C}1$–$\mathcal{C}4$, problem $\textbf{P1}$ is an MINLP, which is generally NP-hard. As a result, solving problem $\textbf{P1}$ efficiently is challenging.
\end{remark}
\vspace{-.5mm}

\vspace{-3mm}
\section{ Lyapunov-Based Decoupling of the Long-Term MINLP} \label{sec:Lyapunov}
\vspace{-.15mm}
Since the decisions regarding channel allocation, power control, and UAV flight altitude adjustment are interdependent over time, it is challenging to satisfy the long-term constraint $\mathcal{C}8$ without knowledge of future realizations of random vehicle positions and time-varying channel conditions. Therefore, in this section, we apply Lyapunov optimization to decouple the multi-stage MINLP problem into per-slot deterministic optimization problems, ensuring the satisfaction of the long-term constraint $\mathcal{C}8$ under stochastic conditions.

Specifically, we introduce a virtual queue $Q(t)$ for the UAV to track the accumulated UAV flight energy cost that exceeds the required threshold. By setting $Q(1)=0$, the virtual queue is updated as follows:
\vspace{-.5mm}
\begin{equation}\label{eq:queue_update}
Q(t+1)=\max \Big\{Q(t)+P(t)\Delta-E^{\text{U}}_{\text{th}},0\Big\}.
\end{equation}
The virtual queue $Q(t)$ is employed to enforce constraint $\mathcal{C}8$ (see Appendix~\ref{app:virtual_queue}). To manage the queue length efficiently, we adopt the quadratic form of the \emph{Lyapunov function}--a well-established tool for simplifying dynamic system analysis~\cite{10736570}. This function is defined as follows:
\vspace{-.5mm}
\begin{equation}\label{eq:Lyapunov_function}
L\Big(Q(t)\Big)=\frac{1}{2}\Big(Q(t)\Big)^2.
\end{equation}
Subsequently, we employ the \emph{conditional Lyapunov drift} to quantify the change in the quadratic Lyapunov function between consecutive time slots, expressed as:
\vspace{-.5mm}
\begin{equation}\label{eq:Lyapunov_drift_function}
\Delta L\Big(Q(t)\Big)=\mathbb{E}\Big\{ L\Big(Q(t+1)\Big)-L\Big(Q(t)\Big)\mid Q(t) \Big\},
\end{equation}
where the expectation accounts for randomness in energy consumption. A high conditional Lyapunov drift value indicates greater likelihood of violating constraint $\mathcal{C}8$, and conversely, a low value suggests higher stability. Finally, to jointly optimize the objective function of problem $\textbf{P1}$ (defined in~\eqref{eq:problem1}) while satisfying the long-term constraint $\mathcal{C}8$, we introduce the \emph{Lyapunov drift-plus-penalty function}:
\vspace{-.5mm}
\begin{equation}\label{eq:Lyapunov_drift_penalty}
D\Big(Q(t)\Big)\!=\!\Delta L\Big(Q(t)\Big)\!-\!V\mathbb{E}\Big\{\frac{1}{M}\! \sum_{m\in\mathcal{M}} R^{\text{U}}_{m}(t)\mid Q(t)\Big\} ,
\end{equation}
where $V > 0$ is an adjustable weight parameter that balances the relative importance between the V2U communication sum rate and long-term UAV energy consumption. We next derive an upper bound on the right-hand side of~\eqref{eq:Lyapunov_drift_penalty}, expressed as (see Appendix~\ref{app:upperbound}):
\vspace{-.5mm}
\begin{align}\label{eq:upper_bound}
&D\Big(Q(t)\Big)\leq \mathbb{E}\Big\{ Q(t)\Big(P(t)\Delta-E^{\text{U}}_{\text{th}}\Big)\mid Q(t)\Big\}\nonumber\\
&-V\mathbb{E}\Big\{\frac{1}{M} \sum_{m\in\mathcal{M}} R^{\text{U}}_{m}(t)\mid Q(t)\Big\}
+\frac{1}{2}\Big(P(t)\Delta-E^{\text{U}}_{\text{th}}\Big)^2.
\end{align}
By omitting the constant which is independent of queue length, minimizing the upper bound in~\eqref{eq:upper_bound} allows us to reformulate the original problem $\textbf{P1}$ as a per-slot deterministic optimization problem $\textbf{P2}$. This problem can be solved in each time slot (details will be provided in Sec.~\ref{sec:D3PG}) without requiring knowledge of future channel states or vehicle mobility patterns, while still satisfying the long-term constraint $\mathcal{C}8$:
\vspace{-1.5mm}
\begin{align}
     \textbf{P2}:\! \min_{\{\bm{x}(t),\bm{p}(t),\Delta H(t)\}}& \!\!  Q(t)\Big(P(t)\Delta\!-\!E^{\text{U}}_{\text{th}}\Big)\!- \!\frac{V}{M}\!\!\sum_{m\in\mathcal{M}}\!\!\! R^{\text{U}}_{m}(t)\label{eq:problem2} \\
     &\quad\quad \text{s.t.} \quad
     \mathcal{C}1-\mathcal{C}7, \nonumber 
\end{align}
where $\bm{x}(t)=\{x_{k,m}(t)\}_{k\in \mathcal{K},m\in\mathcal{M}}$ represents the channel allocation vector for V2V communications reusing the spectrum of V2U communications at time slot $t$, $\bm{p}(t)=\{p_{m}(t),p_k(t)\}_{k\in \mathcal{K},m\in\mathcal{M}}$ denotes the power control vector for the transmitting vehicles of both V2U and V2V communications at time slot $t$, and $\Delta H(t)$ denotes the UAV’s flight altitude adjustment at time slot $t$.

\vspace{-3mm}
\section{Basic Idea of Diffusion Models} \label{sec:diffusion_model}
\vspace{-.15mm}
Prior to introducing our D3PG algorithm, we first present the rationale for combining diffusion models with DRL (specifically, diffusion-based deep deterministic policy gradient). We then describe the adaptation of the diffusion model to generate decisions for channel allocation, power control, and UAV flight altitude adjustment.

\vspace{-3mm}
\subsection{Motivation of Adopting Diffusion Model}  \label{subsec:DDPM_motivation} 
\vspace{-.15mm}
Beyond the limitations of multi-layer perceptrons (MLPs) in conventional DRL approaches (discussed in Sec.~\ref{subsec:DRL_optimization}), our adoption of diffusion models is further motivated by their distinctive compatibility with DRL frameworks. Specifically, in a conventional diffusion model, a user can input a text prompt (e.g., ``an apple on the table'') to guide the model in generating a corresponding image. In our scenario, we conceptualize optimal channel allocation, power control, and UAV flight altitude adjustment as the target \emph{image} to be generated. 

Additionally, integrating diffusion models enables robust decision-making in dynamic environments with CSI feedback delay, as real-time environmental parameters--such as fluctuating channel conditions and stochastic vehicle locations--can be incorporated as conditioning \emph{prompts} during model training. Once trained, the diffusion model can produce optimized decisions adapted to any encountered environmental state~\cite{du2024enhancing}. This dynamic solution-generation capability proves especially advantageous for our dynamic vehicular networks.

% in the reverse process of diffusion models (detailed in Sec.~\ref{subsubsec:reverse_process}), a user-provided text prompt is used to guide the denoising process, generating images that align with the user's preferences (detailed in Sec.~\ref{subsec:DDPM}). 

\vspace{-3mm}
\subsection{Preliminaries of Diffusion Models}  \label{subsec:DDPM}
\vspace{-.15mm}
The denoising diffusion probabilistic model (DDPM)~\cite{ho2020denoising} was initially developed for image generation tasks. In standard DDPM implementation, the training consists of two key stages: the \emph{1) forward process}, which gradually adds noise sampled from a standard Gaussian distribution to an input image over multiple steps until it resembles isotropic Gaussian noise; and the \emph{2) reverse process}, where a neural network learns to systematically remove this noise step-by-step to reconstruct the original image. 

We first combine the optimal channel allocation vector $\bm{x}^*(t)=\{x^*_{k,m}(t)\}_{k\in \mathcal{K},m\in\mathcal{M}}$, the power control vector $\bm{p}^*(t)=\{p^*_{m}(t),p^*_k(t)\}_{k\in \mathcal{K},m\in\mathcal{M}}$, and the UAV’s flight altitude adjustment $\Delta H^*(t)$ at time slot $t$ into a single vector $\bm{\pi}_0(t) = \{\bm{x}^*(t),\bm{p}^*(t),\Delta H^*(t)\}$. This combined vector serves as the optimal solution (i.e., the ``original image'') for our DDPM framework. The forward and reverse processes of this policy are described below.

\begin{figure}[t]
\includegraphics[width=.48\textwidth]{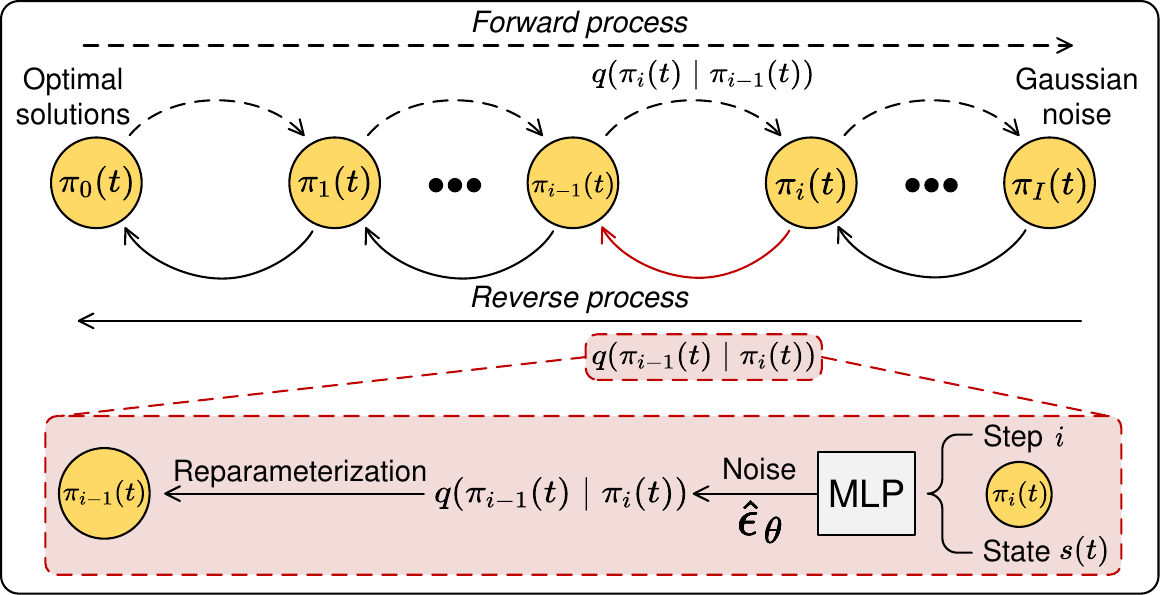}
\centering
\vspace{-1.5mm}
\caption{An illustration of the diffusion model tailored to generate optimal decisions for channel allocation, power control and UAV flight altitude adjustment in time slot $t$.}
\label{fig:DDPM}
\end{figure}

\subsubsection{Forward Process}  \label{subsubsec:forward_process}  
Fig.~\ref{fig:DDPM} illustrates our diffusion model framework for generating optimal channel allocation, power control, and UAV flight altitude adjustment decisions at time slot $t$. Specifically, the forward process follows an $I$-step Markov chain. Beginning with the optimal solution $\bm{\pi}_0(t)$, each step $i$ adds standard Gaussian noise to $\bm{\pi}_{i-1}(t)$, producing $\bm{\pi}_i(t)$. The transition is defined as a normal distribution with a mean of $\sqrt{1-\beta_i}\bm{\pi}_{i-1}(t)$ and a variance of $\beta_i\mathbf{1}$ given by
\vspace{-.5mm}
\begin{align}\label{eq:forward_distribution}
    q\Big(\bm{\pi}_i(t)|\bm{\pi}_{i-1}(t)\Big)= \mathcal{N}\Big(\bm{\pi}_i(t);\sqrt{1-\beta_i}\bm{\pi}_{i-1}(t),\beta_i\mathbf{1}\Big),
\end{align}
where $\beta_i$ is the step-specific diffusion rate~\cite{ho2020denoising}, calculated as $\beta_i=1- e^{-\frac{\beta_{\text{min}}}{I}-\frac{2i-1}{2I^2}(\beta_{\text{max}}-\beta_{\text{min}})}$, with $\beta_{\text{min}}$ and $\beta_{\text{max}}$ being the preset minimum/maximum rates, respectively, and $\mathbf{1}$ denotes the identity matrix.

From~\eqref{eq:forward_distribution}, given that $\bm{\pi}_i(t) \sim \mathcal{N}(\sqrt{1-\beta_i}\bm{\pi}_{i-1}(t),\beta_i\mathbf{1})$, the connection between $\bm{\pi}_{i-1}(t)$ and $\bm{\pi}_i(t)$ can be expressed via reparameterization as follows~\cite{ho2020denoising}:
\vspace{-.5mm}
\begin{equation}\label{eq:forward_update}
    \bm{\pi}_i(t)= \sqrt{1-\beta_i}\bm{\pi}_{i-1}(t)+\sqrt{\beta_i}\bm{\epsilon}_{i-1},
\end{equation}
where $\bm{\epsilon}_{i-1}$ is sampled from $\mathcal{N}(0,\mathbf{1})$. Consequently, using~\eqref{eq:forward_update}, the relationship between $\bm{\pi}_0(t)$ and $\bm{\pi}_i(t)$ at any step $i$ can be derived as:
\vspace{-.5mm}
\begin{align}\label{eq:forward_relationship}
    \bm{\pi}_i(t)=\sqrt{\bar{\varphi}_i}\bm{\pi}_{0}(t)+\sqrt{1-\bar{\varphi}_i}\tilde{\bm{\epsilon}}_i,
\end{align}
where $\bar{\varphi}_i=\prod_{j=1}^{i}\varphi_j$ represents the cumulative product of $\varphi_j$ over the preceding steps $i$, with $\varphi_j=1-\beta_j$, and $\tilde{\bm{\epsilon}}_i \sim \mathcal{N}(0,\mathbf{1})$. 

\vspace{-1mm}
\begin{remark}\label{rem:noforward} Since $\textbf{P2}$ remains an MINLP problem, obtaining the optimal solution $\bm{\pi}_0(t)$--which serves as the \emph{original image} for our DDPM framework--poses significant challenges. Consequently, the forward process is omitted in this work, as indicated by the dotted lines in Fig.~\ref{fig:DDPM}. Instead, the forward process here primarily defines the mathematical relationship between $\bm{\pi}_0(t)$ and $\bm{\pi}_i(t)$, a necessary foundation for the subsequent reverse process.
\end{remark}
\vspace{-.5mm}

\subsubsection{Reverse Process}  \label{subsubsec:reverse_process} 
From~\eqref{eq:forward_relationship}, we note that as $I$ becomes sufficiently large, $\bm{\pi}_I(t)$ converges to standard Gaussian noise. Therefore, in the reverse process, we initialize with $\bm{\pi}_I(t) \sim \mathcal{N}(0, \mathbf{1})$ and employ an MLP-based denoiser $\eta_{\bm{\theta}}$ (parameterized by $\bm{\theta}$) that accepts three inputs: the current decision $\bm{\pi}_{i}(t)$, the step index $i$, and the system state $\bm{s}(t)$ (defined later in Sec. \ref{subsubsec:D3PG_MDP}). Specifically, the denoiser predicts the noise component to be subtracted, thereby recovering $\bm{\pi}_{i-1}(t)$. This transition follows a Gaussian distribution~\cite{ho2020denoising}:
\vspace{-.5mm}
\begin{align}\label{eq:reverse_distribution}
q\Big(\bm{\pi}_{i-1}(t)|\bm{\pi}_i(t)\Big)= \mathcal{N}\Big(\bm{\pi}_{i-1}(t);\bm{\mu}_i(t),\bar{\beta}_i\mathbf{1}\Big),
\end{align}
where $\bar{\beta}_i=\frac{1-\bar{\varphi}_{i-1}}{1-\bar{\varphi}_i}\beta_i$. The mean $\bm{\mu}_i(t)$ is derived through Bayesian inference:
\vspace{-.5mm}
\begin{align}\label{eq:reverse_original_mean}
\bm{\mu}_i(t)= \frac{\sqrt{\varphi_i}(1-\bar{\varphi}_{i-1})}{1-\bar{\varphi}_i}\bm{\pi}_{i}(t)+\frac{\sqrt{\bar{\varphi}_{i-1}}\beta_i}{1-\bar{\varphi}_i}\bm{\pi}_{0}(t).
\end{align}

Next, by substituting~\eqref{eq:forward_relationship} into~\eqref{eq:reverse_original_mean}, we eliminate the dependence on $\bm{\pi}_{0}(t)$ and reformulate the mean $\bm{\mu}_i(t)$ as:
\vspace{-.5mm}
\begin{align}\label{eq:reverse_mean}
&\bm{\mu}_{\bm{\theta}}(\bm{\pi}_i(t),i,\bm{s}(t))\nonumber\\
&=\frac{1}{\sqrt{\varphi_i}}\Big[\bm{\pi}_i(t)-\frac{1-\varphi_i}{\sqrt{1-\bar{\varphi}_i}}\bm{\hat{\epsilon}}_{\bm{\theta}}(\bm{\pi}_i(t),i,\bm{s}(t))\Big],
\end{align}
where $\bm{\hat{\epsilon}}_{\bm{\theta}}(\bm{\pi}_i(t),i,\bm{s}(t))$ denotes the noise estimate produced by the denoiser $\eta_{\bm{\theta}}$ at step $i$. 

Finally, from~\eqref{eq:reverse_distribution}, we derive the transition between consecutive states via reparameterization:
\vspace{-.5mm}
\begin{equation}\label{eq:reverse_update}
    \bm{\pi}_{i-1}(t)= \bm{\mu}_{\bm{\theta}}(\bm{\pi}_i(t),i,\bm{s}(t))+\sqrt{\bar{\beta}_i}\bar{\bm{\epsilon}}_i,
\end{equation}
with $\bar{\bm{\epsilon}}_i \sim \mathcal{N}(0,\mathbf{1})$. In our framework, the denoiser $\eta_{\bm{\theta}}$ serves as the optimal policy network. Through iterative application of~\eqref{eq:reverse_update} over $I$ steps (detailed in Algorithm~\ref{algo：D3PG}), we recover the optimal decisions $\bm{\pi}_0(t) = \{\bm{x}^*(t),\bm{p}^*(t),\Delta H^*(t)\}$ for channel allocation, power control, and UAV flight altitude adjustment.

\vspace{-1mm}
\begin{remark}\label{rem:noforward}
In standard DDPM implementations, the training objective involves minimizing the mean squared error between the forward process noise $\bm{\epsilon}_i$ (sample from $\mathcal{N}(0,\mathbf{1})$) and the denoiser's predicted noise $\bm{\hat{\epsilon}}_{\bm{\theta}}$ at each reverse step. However, our approach differs as we omit the explicit forward process. Instead of relying on the optimal solution $\bm{\pi}_0(t)$, which serves as the original image, we optimize the reverse process through an exploration-based learning strategy. As described in Sec.~\ref{subsec:D3PG_architecture}, this is done by directly minimizing the objective function in~\eqref{eq:problem2}.
\end{remark}
\vspace{-.5mm}

\vspace{-3mm}
\section{Diffusion-Based Deep Deterministic Policy Gradient Algorithm}\label{sec:D3PG}
\vspace{-.15mm}
Henceforth, we first define of the Markov decision process (MDP) elements, followed by an overview of the D3PG algorithm’s architecture. Then, we conduct a comprehensive analysis of its computational complexity.

\subsection{MDP Elements in the D3PG Algorithm} \label{subsubsec:D3PG_MDP}
The sequential decision-making nature of problem $\textbf{P2}$ can be captured via an MDP, which includes the \emph{state space}, \emph{action space}, and \emph{reward function}, as described below.

\begin{itemize}[leftmargin=4mm]
\item \emph{State Space:} In time slot $t$, the DRL agent acting as the central controller (e.g.,  UAV) observes the state $\bm{s}(t)$ to gather environmental information. This state consists of $MK+2K+M+1$ elements, defined as:
\vspace{-.5mm}
\begin{align}\label{eq:D3PG_state}
    \bm{s}(t)=\Big\{\bm{h}^{\text{U}}(t),\bm{h}^{\text{V}}(t),Q(t)\Big\},
\end{align}
where $\bm{h}^{\text{U}}(t)=\{h^{\text{U}}_{m}(t),h^{\text{U}}_{k}(t)\}_{m\in\mathcal{M},k\in \mathcal{K}}$ represents the channel gains of all V2U communications, including interference signals, in time slot $t$, while $\bm{h}^{\text{V}}(t)=\{h^{\text{V}}_{m,k}(t),h^{\text{V}}_{k}(t)\}_{m\in\mathcal{M},k\in \mathcal{K}}$ captures the channel gains of all V2V communications, also incorporating interference signals, in time slot $t$, and $Q(t)$ indicates the current virtual queue status in time slot $t$.

\item \emph{Action Space:} In time slot $t$, the action space comprises decisions for channel allocation, power control and UAV flight altitude adjustment, containing $2K+M+1$ elements, expressed as
\vspace{-.5mm}
\begin{align}\label{eq:D3PG_action}
    \bm{a}(t)=\Big\{\bm{x}(t),\bm{p}(t),\Delta H(t)\Big\},
\end{align}
where $\bm{x}(t)=\{x_{k,m}(t)\}_{k\in \mathcal{K},m\in\mathcal{M}}$ denotes the channel allocation vector for V2V communications, $\bm{p}(t)=\{p_{m}(t),p_k(t)\}_{k\in \mathcal{K},m\in\mathcal{M}}$ represents the power control vector for transmitting vehicles in both V2U and V2V communications during time slot $t$, and $\Delta H(t)$ indicates the UAV's flight altitude adjustment at time slot $t$. Note that the initial actions produced by the D3PG algorithm are $\tilde{\bm{a}}(t)=\{\tilde{\bm{x}}(t),\tilde{\bm{p}}(t),\Delta\tilde{H}(t)\}$, with elements normalized to the range $[-1,1]$. We subsequently employ an action amender~\cite{liu2024improved} to guarantee that that all actions $\bm{a}(t)=\{\bm{x}(t),\bm{p}(t),\Delta H(t)\}$ comply with the constraints specified in problem $\textbf{P2}$. Specifically, to satisfy constraints $\mathcal{C}1$, $\mathcal{C}5$, and $\mathcal{C}6$, the channel allocation decision is represented by a $K \times M$ matrix, where the values in each row indicate the preference scores for assigning each channel to the corresponding link. For each V2V link, the channel with the highest score in its row is selected as the final channel assignment. Then, to satisfy constraints $\mathcal{C}2$ and $\mathcal{C}3$, the power control decisions are normalized as $p_{m}(t)=\frac{\tilde{p}_{m}(t)+1}{2}p_{\text{max}}$ and $p_{k}(t)=\frac{\tilde{p}_{k}(t)+1}{2}p_{\text{max}}$, respectively. Furthermore, to satisfy constraint $\mathcal{C}4$, the UAV’s flight altitude adjustment decision is scaled to $\Delta H(t)=\Delta\tilde{H}(t)\times\Delta H_{\text{max}}$, where $\Delta H_{\text{max}}$ denotes the maximum altitude the UAV can adjust in each time slot.

\item \emph{Reward Function:} After executing action $\bm{a}(t)$ based on state $\bm{s}(t)$, the environment returns a reward $r(t)$ as feedback. This reward is defined as the negative value of the objective function in~\eqref{eq:problem2}, since the D3PG algorithm aims to maximize the reward during training, as follows:
\vspace{-.5mm}
\begin{align}\label{eq:D3PG_reward}
    r(t)=\frac{V}{M}\sum_{m\in\mathcal{M}} R^{\text{U}}_{m}(t)-Q(t)\Big(P(t)\Delta-E^{\text{U}}_{\text{th}}\Big)\nonumber\\- \mathbb{I}\Big\{\text{Pr}\{\gamma^{\text{V}}_k(t) < \gamma^{\text{V}}_{\text{th}} \} > \text{Pr}^{\text{V}}_{\text{th}}\Big\}\Gamma^{\text{pen}},
\end{align}
where $\mathbb{I}\{\cdot\}$ denotes an indicator function that equals 1 when the condition is satisfied and 0 otherwise. $\Gamma^{\text{pen}}$ represents a constant penalty term to prevent the agent from violating constraint $\mathcal{C}7$.
\end{itemize}

\begin{figure}[t]
\includegraphics[width=.48\textwidth]{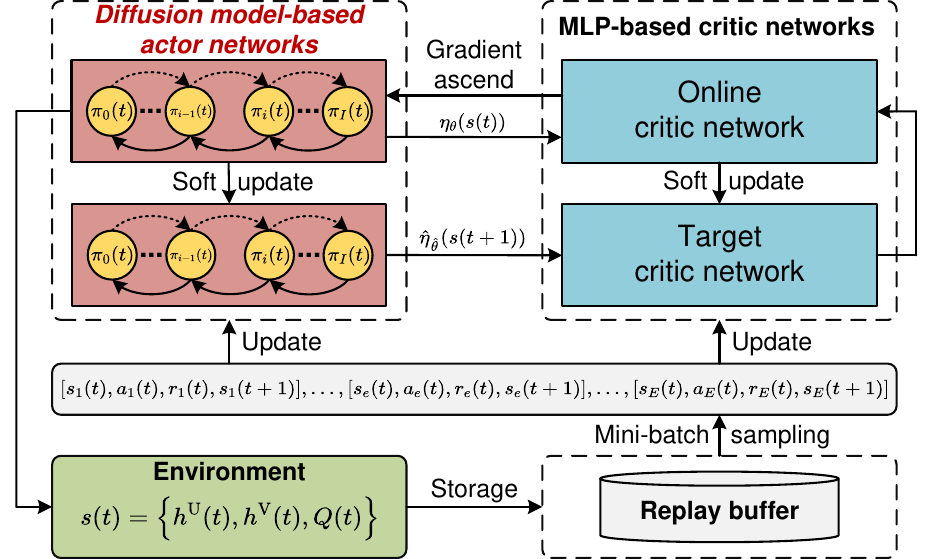}
\centering
\vspace{-1.5mm}
\caption{The overall architecture of the D3PG algorithm.}
\label{fig:GDM_DDPG}
\end{figure}

\subsection{Architecture of the D3PG Algorithm}  \label{subsec:D3PG_architecture}
The architecture of D3PG is illustrated in Fig.~\ref{fig:GDM_DDPG}, consisting of an \emph{online diffusion model-based actor network} responsible for action generation and an \emph{online critic network} for action evaluating. To mitigate training instability, two \emph{target networks} are incorporated. Additionally, a \emph{replay buffer} is used to reduce sample correlation through random sampling.

\begin{itemize}[leftmargin=4mm]
\item \emph{Diffusion Model-Based Actor Network:} In D3PG, the actor network $\eta_{\bm{\theta}}$, parameterized by $\bm{\theta}$, is built around the denoiser from the diffusion model introduced in Sec.~\ref{sec:diffusion_model}, rather than a traditional MLP. To enhance training stability, a target actor network $\hat{\eta}_{\bm{\hat{\theta}}}$, sharing the same architecture as $\eta_{\bm{\theta}}$ and parameterized by $\bm{\hat{\theta}}$, is also employed.

\item \emph{Critic Network:} The critic network $\mathbb{Q}_{\bm{\phi}}$, parameterized by $\bm{\phi}$, is implemented as an MLP that takes the state $\bm{s}(t)$ and action $\bm{a}(t)$ as inputs and outputs the Q-value $\mathbb{Q}_{\bm{\phi}}(\bm{s}(t),\bm{a}(t))$. This Q-value quantifies the expected quality of the state-action pair, where a higher value suggests a greater likelihood of achieving a higher reward. To further improve training stability, a target critic network $\hat{\mathbb{Q}}_{\bm{\hat{\phi}}}$, with parameters $\bm{\hat{\phi}}$, and the same architecture, is also employed.

\item \emph{Replay Buffer:} During training, a replay buffer $\mathcal{E}$ is utilized to store transition tuples. At each time slot $t$, D3PG stores the tuple $[\bm{s}(t), \bm{a}(t), r(t), \bm{s}(t+1)]$ in $\mathcal{E}$, where it is retained for future sampling to support policy learning.

% For simplicity, we denote the state, action, and reward in the current time slot by $\bm{s}_t$, $\bm{a}_t$ and $r_t$, respectively, while $\bm{s}^{\prime}_t$ represents the state in the next time slot. 

\item \emph{Policy Improvement:} After a certain amount of exploration, a mini-batch of $E$ samples $\{[\bm{s}_e(t),\bm{a}_e(t),r_e(t),\bm{s}_{e}(t+1)]\}_{e=1}^{E}$ is randomly drawn from the replay buffer $\mathcal{E}$ to update both the critic and actor networks. For the critic network $\mathbb{Q}_{\bm{\phi}}$ in particular, the update aims to minimize the temporal difference (TD) error between the target Q-value $y_e(t)$ and the predicted Q-value $\mathbb{Q}_{\bm{\phi}}(\bm{s}_e(t),\bm{a}_e(t))$, as defined by
\vspace{-.5mm}
\begin{align}\label{eq:D3PG_critic}
    \text{TD}^{\text{error}}=\frac{1}{E}\sum_{e=1}^E\Big[(y_e(t)-\mathbb{Q}_{\bm{\phi}}(\bm{s}_e(t),\bm{a}_e(t))^2\Big],
\end{align}
where $y_e(t)=r_e(t)+\omega \hat{\mathbb{Q}}_{\bm{\hat{\phi}}}(\bm{s}_{e}(t+1),\hat{\eta}_{\hat{\bm{\theta}}}(\bm{s}_{e}(t+1)))$. In this expression, $e$ indexes the $e$-th transition tuple sampled from the replay buffer $\mathcal{E}$, and $\omega$ is the discount factor that weights future rewards. Additionally, the target Q-value $y_e(t)$ is calculated using the target critic network $\hat{\mathbb{Q}}_{\bm{\hat{\phi}}}$. Specifically, this network receives the next state $\bm{s}_{e}(t+1)$ and the corresponding next action $\hat{\eta}_{\hat{\bm{\theta}}}(\bm{s}_{e}(t+1))$, generated by the target actor network, as inputs and outputs the associated target Q-value. The estimation accuracy of $\mathbb{Q}_{\bm{\phi}}$ is then improved by iteratively minimizing the loss in~\eqref{eq:D3PG_critic} using a standard optimizer, such as Adam~\cite{kingma2014adam}, as follows:
\vspace{-.5mm}
\begin{align}\label{eq:D3PG_criticnet_update}
    \bm{\phi} \leftarrow \bm{\phi} - \sigma^{\text{critic}}\text{TD}^{\text{error}},
\end{align}
where $\sigma^{\text{critic}}$ denotes the learning rate of the critic network. In parallel, the actor network $\eta_{\bm{\theta}}$ is updated using the sample policy gradient:
\vspace{-.5mm}
\begin{align}\label{eq:D3PG_actor}
    &\nabla_{{\bm{\theta}}}J= \frac{1}{E}\sum_{e=1}^E \Big\{\nonumber \\
    &\nabla_{\bm{a}}\mathbb{Q}_{\bm{\phi}}(\bm{s}_e(t),\bm{a})\mid_{\bm{a}=\eta_{\bm{\theta}}(\bm{s}_e(t))}\nabla_{\bm{\theta}}\eta_{\bm{\theta}}(\bm{s}_e(t))\Big\},
\end{align}
where the actor network $\eta_{\bm{\theta}}$ is optimized via gradient ascent based on~\eqref{eq:D3PG_actor} to maximize the cumulative reward defined in~\eqref{eq:D3PG_reward}. This is typically performed using a standard optimizer such as Adam~\cite{kingma2014adam}, as follows: 
\vspace{-.5mm}
\begin{align}\label{eq:D3PG_actornet_update}
    \bm{\theta} \leftarrow \bm{\theta} + \sigma^{\text{actor}}\nabla_{{\bm{\theta}}}J,
\end{align}
where $\sigma^{\text{actor}}$ is the learning rate for the actor network. To ensure stable training, the parameters of the target networks are updated gradually, promoting smooth changes in the learned policy and Q-value estimates over time. This is achieved through soft updates as follows: 
\vspace{-.5mm}
\begin{align}
    &\hat{\bm{\theta}}\leftarrow \tau\bm{\theta}+(1-\tau)\hat{\bm{\theta}}, \label{eq:D3PG_target_actor}\\
    &\hat{\bm{\phi}} \leftarrow \tau\bm{\phi}+(1-\tau)\hat{\bm{\phi}}, \label{eq:D3PG_target_critic}
\end{align}
where $\tau \in (0,1]$ denotes the update rate of the target networks.

\end{itemize}

\begin{algorithm} [!t]
\footnotesize
  \SetAlgoLined
  \SetKwData{Left}{left}\SetKwData{This}{this}\SetKwData{Up}{up}
  \SetKwFunction{Union}{Union}\SetKwFunction{FindCompress}{FindCompress}
  \SetKwInOut{Input}{input}\SetKwInOut{Output}{output}
  %\Input{A bitmap $Im$ of size $w\times l$}
  %\Output{A partition of the bitmap}
  \textbf{Input:} Initialize the network parameters $\bm{\theta}$ and $\bm{\phi}$, and set all hyperparameters, including the number of learning episodes $S$, discount factor $\omega$, penalty term $\Gamma^{\text{pen}}$, learning rates $\sigma^{\text{critic}}$ and $\sigma^{\text{actor}}$, replay buffer $\mathcal{E}$, and target network update rate $\tau$.

  \textbf{Output:} The optimal channel allocation, power control, and UAV flight altitude adjustment decisions.
  \BlankLine
  
  \For{$episode=1$ \KwTo $S$}{
        
        \For{$t=1$ \KwTo $T$}{ 
            Observe the environment to obtain $\bm{s}(t)$ according to~\eqref{eq:D3PG_state} and initialize a distribution $\bm{\pi}_I(t)\sim\mathcal{N}(0,\mathbf{1})$.
      
            \For{$i=I$ \KwTo $0$} {
            Use a MLP-based denoiser $\eta_{\bm{\theta}}$ (parameterized by $\bm{\theta}$) to infer the noise $\bm{\hat{\epsilon}}_{\bm{\theta}}(\bm{\pi}_i(t),i,\bm{s}(t))$.
        
            Calculate the mean $\bm{\mu}_{\bm{\theta}}(\bm{\pi}_i(t),i,\bm{s}(t))$ and the distribution $q(\bm{\pi}_{i-1}(t)|\bm{\pi}_i(t))$ by~\eqref{eq:reverse_mean} and~\eqref{eq:reverse_distribution}, respectively. 

            Calculate the distribution $\bm{\pi}_{i-1}(t)$ using the reparameterization technique~\eqref{eq:reverse_update}.
		  }
      
        Obtain the optimal channel allocation, power control, and UAV flight altitude adjustment decisions as $\bm{\pi}_0(t) = \{\bm{x}^*(t),\bm{p}^*(t),\Delta H^*(t)\}$.
        
        Receive the reward $r(t)$ according to~\eqref{eq:D3PG_reward} and transition to the next state $\bm{s}(t+1)$.
      
        Store $[\bm{s}(t),\bm{a}(t),r(t),\bm{s}(t+1)]$ into $\mathcal{E}$. 
        
        Randomly sample a batch of $E$ transitions $\{[\bm{s}_e(t),\bm{a}_e(t),r_e(t),\bm{s}_{e}(t+1)]\}_{e=1}^E$ from $\mathcal{E}$.

        Update the online networks' parameters $\bm{\phi}$ and $\bm{\theta}$ by~\eqref{eq:D3PG_critic} and~\eqref{eq:D3PG_actor}, respectively.

        Update the target networks' parameters $\hat{\bm{\theta}}$ and $\bm{\hat{\phi}}$ by~\eqref{eq:D3PG_target_actor} and~\eqref{eq:D3PG_target_critic}, respectively.
        }
    }
  \caption{D3PG Algorithm.}\label{algo：D3PG}
\end{algorithm}\DecMargin{1em}

\vspace{-3mm}
\subsection{D3PG Algorithm and Complexity Analysis}
\vspace{-.15mm}
Algorithm~\ref{algo：D3PG} presents the pseudocode of the proposed D3PG algorithm. Note that since the training process of learning-based methods can be performed in a cloud data center with ample computational resources~\cite{10736570}, our analysis of the D3PG algorithm’s computational complexity primarily focuses on its execution phase (i.e., the inference process). Assume that the denoiser in D3PG consists of $L$ fully connected layers, each containing $n$ neurons. A single forward pass through the denoiser then requires $\mathcal{O}(Ln^2)$ multiply–accumulate operations. Since the diffusion model performs $I$ denoising steps to generate the action at each time slot, the per-slot complexity is $\mathcal{O}(ILn^2)$. Finally, across $S$ episodes, each containing $T$ time slots, the total computational complexity of the proposed D3PG algorithm is $\mathcal{O}(STILn^2)$, which is linear in the number of episodes ($S$), time slots per episode ($T$), and denoising steps ($I$), and quadratic in the width of the network’s hidden layers ($n$).

\vspace{-3mm}
\section{Performance Evaluation}  \label{sec:simulation} 
\vspace{-.15mm}

In this section, we first present the simulation parameter settings and then evaluate the performance of the proposed D3PG by comparing it with three benchmark solutions.

\begin{figure}[htb]
    \centering
    \begin{subfigure}[b]{0.24\textwidth}  % [b]是基线对齐，0.45\textwidth是子图宽度
        \includegraphics[width=\textwidth]{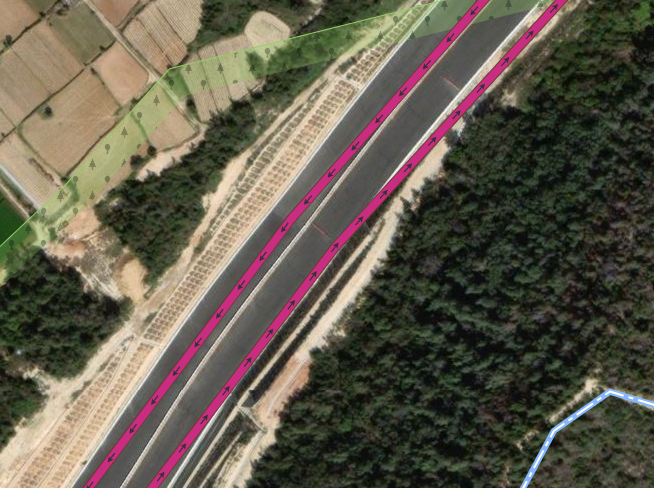}  % 替换为你的图片文件
        \caption{Real-world traffic region.}
        \label{fig:real-world traffic}
    \end{subfigure}
    \hfill  % 在两个子图之间添加一些水平空间
    \begin{subfigure}[b]{0.216\textwidth}
        \includegraphics[width=\textwidth]{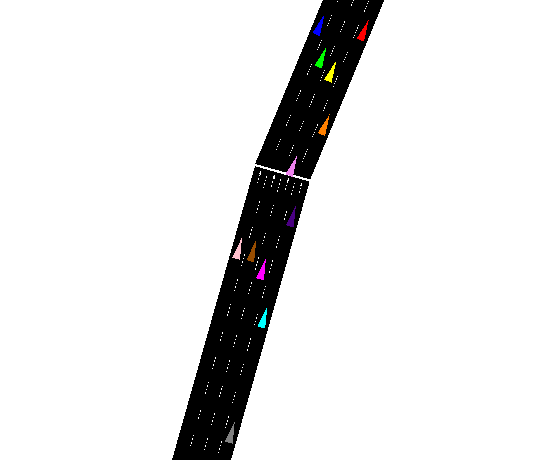}  % 替换为你的图片文件
        \caption{Import moving vehicles.}
        \label{fig:moving_vehicles}
    \end{subfigure}
    \vspace{-1.5mm}
    \caption{Vehicular network visualization.}
    \label{fig:vehicular_networks}
\end{figure}

\vspace{-3mm}
\subsection{Simulation Settings}
\vspace{-.15mm}
\subsubsection{Network Layout} We consider a real-world, one-way highway in Xiamen, China, with a length of 2 km, as shown in Fig.\ref{fig:real-world traffic}, based on data obtained from OpenStreetMap\cite{haklay2008openstreetmap}. The SUMO simulator~\cite{lopez2018microscopic} is then used to generate moving vehicles,\footnote{To enhance road capacity, vehicles travel in platoons at an average speed of 50 km/h, ensuring uninterrupted V2U and V2V communications throughout the simulation~\cite{he2023aerial}. SUMO is also used to record the positions of vehicles at different time slots, thereby capturing the dynamic characteristics of the vehicular network.} resulting in a realistic vehicular network illustrated in Fig.\ref{fig:moving_vehicles}. Additionally, a standalone UAV travels at a constant speed of 50 km/h, following the ground vehicles to serve as an aerial base station and provide communication services. The main simulation parameters are summarized in Table~\ref{table2}.

\subsubsection{Algorithm Layout} We implement D3PG using PyTorch 2.7.0 and Python 3.12.4 on a platform equipped with an Intel Core i7-7700 CPU. For the diffusion model, the denoiser is implemented using three fully connected (FC) hidden layers. The critic networks in D3PG are similarly constructed with three FC hidden layers. We use the Adam optimizer with learning rates of $\sigma^{\text{critic}} = 10^{-5}$ and $\sigma^{\text{actor}} = 3\times10^{-6}$ for the critic and actor networks, respectively. The ReLU activation function is applied to each hidden layer, while a tanh activation function is used in the denoiser’s output layer to constrain the action range.

\begin{table}[!t]
\vspace{-2mm}
\centering
\footnotesize
\caption{Parameters used in simulations ~\cite{he2023aerial, zhang2024performance, dai2023uav, cai2022resource, yang2022online}.}
\label{table2}
\rowcolors{1}{white}{gray!20}
% \rowcolors{1}{lightblue!30}{white}
\begin{tabular}{|l|c|}
\hline
% \rowcolor{lightgray} % 设置这一行的背景色为浅灰色
\textbf{Parameter}                    & \textbf{Value}   \\ \hline \hline
Number of time slots ($T$)                  & 100 seconds              \\ \hline
Duration of each time slot ($\Delta$)          & 1 second        \\ \hline
Number of V2U communications ($M$)                   & 10                \\ \hline
Maximum transmission power ($p_{\text{max}}$)   & 23 dBm           \\ \hline
Range of UAV altitude ($[H_{\text{min}},H_{\text{max}}]$)                  & [50, 200] m                \\ \hline
Maximum altitude the UAV can adjust ($\Delta H_{\text{max}}$)                  & 5 m                \\ \hline
Bandwidth of each channel ($B$)  & 2 MHz           \\ \hline
Noise power spectral density ($N_0$)                  & -174 dBm/Hz        \\ \hline
Carrier frequency ($f_c$)                  & 5.9 GHz        \\ \hline
Additional losses under LoS ($\alpha_{\text{LoS}}$)                          & 1 dB      \\ \hline
Additional losses under NLoS ($\alpha_{\text{NLoS}}$)                          & 20 dB        \\ \hline
SINR requirement of V2V links ($\gamma^{\text{V}}_{\text{th}}$)                  & 10 dB                \\ \hline
Tolerable outage probability ($\text{Pr}^{\text{V}}_{\text{th}}$)                  & 1.0 \%                \\ \hline
Maximum allowed operational power ($E^{\text{U}}_{\text{th}}$)                  & 120 J                \\ \hline
Number of episodes ($S$)                          & 500        \\\hline
D3PG's reward penalty ($\Gamma^{\text{pen}}$)                  & 10          \\ \hline
D3PG's reward discount factor ($\omega$)                  & 0.99          \\ \hline
D3PG's target network update rate ($\tau$)                          & 0.005        \\ \hline
Parameters for environment ($a,b$)                          & 12.08, 0.11       \\ \hline
Weight of UAV ($G$)                          & 20 Newton        \\ \hline
Blade angular velocity ($\Omega$)                          & 300 radians/second       \\ \hline
Rotor radius ($r$)                          & 0.4 meter        \\ \hline
Air density ($\rho$)                          & 1.225 kg/$\text{m}^3$       \\ \hline
Rotor solidity ($s_r$)                          & 0.05 $\text{m}^3$        \\ \hline
Rotor disc area ($A_r$)                          & 0.503 $\text{m}^2$        \\ \hline
Induced velocity for rotor ($v_0$)                          & 4.03 meter/second        \\ \hline
Fuselage drag ratio ($d_0$)                          & 0.3        \\ \hline
Blade profile power in hovering ($P_0$)                          & 79.86 W        \\ \hline
Induced power in hovering ($P_1$)                          & 88.63 W        \\ \hline
\end{tabular}
\end{table}

\vspace{-3mm}
\subsection{Benchmark Solutions}
\vspace{-.15mm}
To demonstrate the effectiveness of the proposed D3PG algorithm, we have relied on three benchmark solutions:

\begin{itemize}[leftmargin=4.5mm] 
    \item \emph{DDPG:} Channel allocation, power control, and the UAV flight altitude adjustment are optimized by the DDPG algorithm~\cite{lillicrap2015continuous}. Unlike our proposed D3PG, which incorporates a diffusion model, DDPG employs an MLP-based actor network to make decisions. This baseline is used to highlight the significant performance gains achieved by leveraging the diffusion model in our approach.
    \item \emph{D3PG without considering CSI feedback delay (D3PG-WCSI):} Channel allocation, power control, and UAV flight altitude adjustment are optimized using the same strategy as in D3PG; however, the corresponding V2V communication model formulation does not account for CSI feedback delay. This baseline is designed to highlight the Doppler effect caused by the high mobility in vehicular networks and to demonstrate the necessity of considering CSI feedback delay. 
    \item \emph{Hungarian and DDQN-based resource allocation algorithm (H-DDQN)~\cite{qi2022energy}:} Channel allocation is optimized using the Hungarian algorithm, while power control and UAV flight altitude adjustment are optimized using the DDQN algorithm~\cite{van2016deep}. As DDQN is a value-based learning algorithm, power control levels and UAV flight altitudes are discretized into predefined values to fit its framework.
\end{itemize}

% \begin{figure}[t!]
%     \centering
%     \includegraphics[width=.4\textwidth]{Result_Figures/T2DRL_convergence_different_steps.pdf}
%     \vspace{-3mm}
%     \caption{Impact of actor learning rate on the reward in D3PG (number of V2V links $K=$, Lyapunov weight $V=$, and CSI feedback delay $T_{\text{delay}}=$).}
%     \label{fig:D3PG_convergency_different_LR}
%     \vspace{-3mm}
% \end{figure}

\vspace{-3mm}
\subsection{Simulation Results}
\vspace{-.15mm}
To eliminate the influence of randomness and ensure a fair comparison, we run each algorithm five times under different environmental settings (i.e., using five different random seeds) and use the average results to generate the following figures.

% \subsubsection{Effect of the Value of Actor Learning Rate $\sigma^{\text{actor}}$} In Fig.~\ref{fig:D3PG_convergency_different_LR}, we illustrate the convergence behavior of the D3PG algorithm under different actor learning rates. The results show that the converged reward initially improves as the learning rate increases, but then declines beyond a certain threshold. This is because a moderate increase in the learning rate helps the D3PG agent escape local optima, whereas an excessively high learning rate may cause the actor network to overfit to inaccurate value estimates from the critic network, leading to reward fluctuations or degraded performance. Based on these observations, we set the actor network's learning rate in D3PG to $\sigma^{\text{actor}}=$ for comparisons with benchmark solutions in the subsequent experiments.

\subsubsection{Effect of the Value of Denoising Step $I$} 
\begin{figure}[t!]
    \centering
    \includegraphics[width=.4\textwidth]{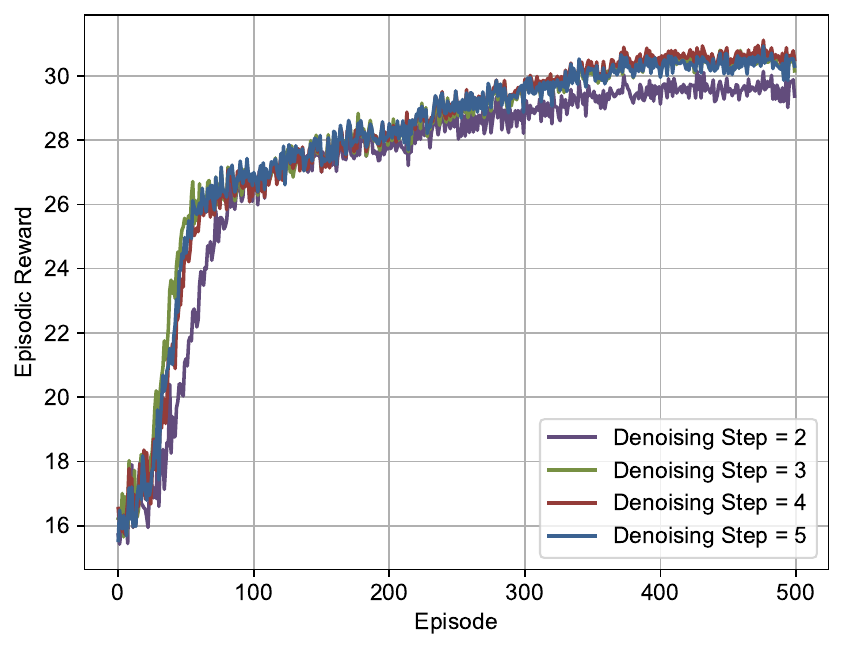}
    \vspace{-3mm}
    \caption{Impact of denoising step on the reward in D3PG (the number of V2V links $K=10$, Lyapunov weight $V=100$, and CSI feedback delay $T_{\text{delay}}=10$ms).}
    \label{fig:D3PG_convergency_different_DStep}
    \vspace{-3mm}
\end{figure}

In Fig.~\ref{fig:D3PG_convergency_different_DStep}, we present the convergence behavior of the D3PG algorithm under varying numbers of denoising steps $I$ in the diffusion model, which directly influences the action sampling process. The results show that the converged reward initially improves with an increasing number of denoising steps, but begins to decline beyond a certain point. This is because a moderate number of denoising steps stabilizes training and allows the diffusion model to capture more generalizable features. However, an excessive number of steps may over-smooth the output, removing useful signal components and ultimately degrading performance. Based on this observation, we set the number of denoising steps in D3PG to $I=4$ for comparison with benchmark solutions in the subsequent experiments.

\begin{figure}[t!]
    \centering
    \includegraphics[width=.4\textwidth]{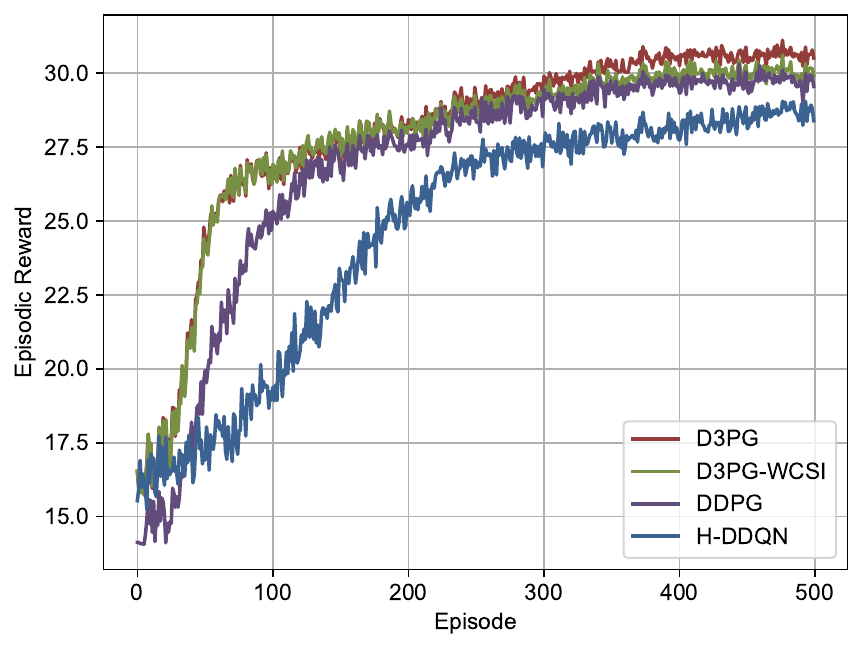}
    \vspace{-3mm}
    \caption{Comparison of reward curves among different algorithms (the number of V2V links $K=10$, Lyapunov weight $V=100$, and CSI feedback delay $T_{\text{delay}}=10$ms).}
    \label{fig:different_algo_convergency}
\end{figure}

\subsubsection{Convergence Performance} In Fig.~\ref{fig:different_algo_convergency}, we depict the convergence behavior of four different algorithms as the number of training episodes increases. The results show that the proposed D3PG achieves the highest episodic reward among all methods. Specifically, this superiority stems from the use of a diffusion model in D3PG's actor network, in contrast to the conventional MLP used in DDPG, which generates actions through a single forward pass and often suffers from limited exploration capability and susceptibility to local optima—particularly in complex, high-dimensional action spaces. In contrast, diffusion model-based actor networks generate actions through a step-wise denoising process, allowing for iterative refinement and stochastic exploration. This iterative nature enables policies to explore the solution space more effectively and avoid premature convergence. This underscores the effectiveness of diffusion-based policy representation in capturing optimal actions in complex environments.

In comparison, D3PG-WCSI performs worse than D3PG. The key difference lies in how CSI feedback delay is handled. D3PG-WCSI neglects the impact of delayed CSI in the agent’s state observations, leading the agent to learn policies based on outdated or inaccurate V2V communication states. However, during reward computation, the V2U communication sum rate is calculated using the true delayed CSI, resulting in a mismatch between the observed state and the actual environment dynamics. This discrepancy leads to suboptimal learning and degraded performance, highlighting the importance of explicitly modeling CSI feedback delay in the decision-making process.

The H-DDQN algorithm exhibits the lowest performance, primarily due to its value-based nature, which necessitates discretizing both transmission power and UAV altitude into finite levels. This discretization reduces the granularity of the action space, limiting the agent’s ability to fine-tune its control decisions. In contrast, policy-based methods such as DDPG and D3PG operate in continuous action spaces, allowing them to learn more precise and adaptable strategies.

% \begin{figure*}[t!]
%     \centering
%     \begin{minipage}[b]{0.32\textwidth}
%         \centering
%         \includegraphics[width=\linewidth]{Result_Figures/DDPG_D3PG_convergence.pdf}
%         \vspace{-3.5mm}
%         \caption{Task queue length of different CVs.}
%         \label{fig:cv_queue_stability}
%     \end{minipage}
%     \hfill
%     \begin{minipage}[b]{0.32\textwidth}
%         \centering
%         \includegraphics[width=\linewidth]{Result_Figures/DDPG_D3PG_convergence.pdf}
%         \vspace{-3.5mm}
%         \caption{Task queue length of different DNNs.}
%         \label{fig:dnn_queue_stability}
%     \end{minipage}
%     \hfill
%     \begin{minipage}[b]{0.32\textwidth}
%         \centering
%         \includegraphics[width=\linewidth]{Result_Figures/DDPG_D3PG_convergence.pdf}
%         \vspace{-3.5mm}
%         \caption{Task queue length of different SVs.}
%         \label{fig:sv_queue_stability}
%     \end{minipage}
% \end{figure*}
\begin{figure}[t!]
    \centering
    \includegraphics[width=.4\textwidth]{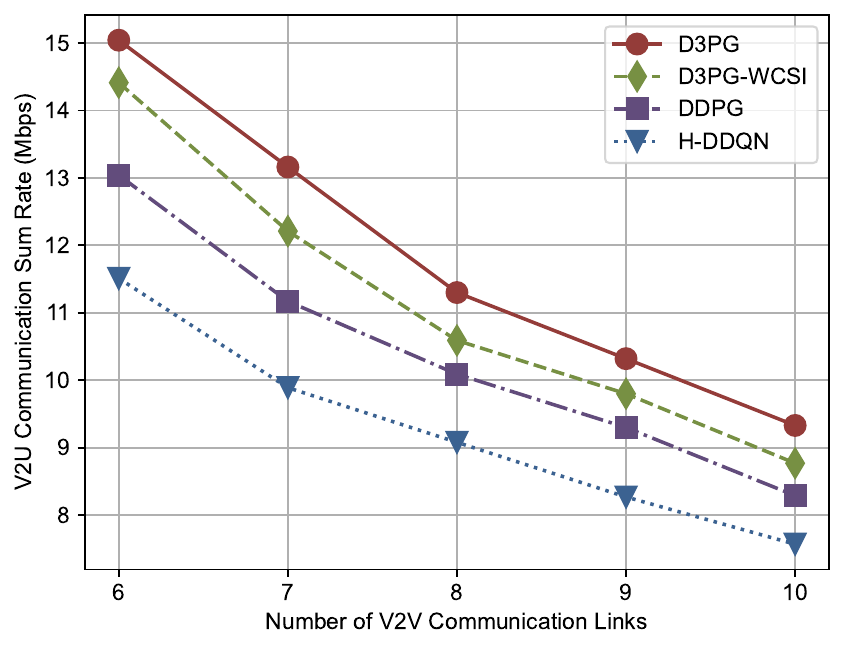}
    \vspace{-3mm}
    \caption{Impact of the number of V2V links on the V2U communication sum rate in~\eqref{eq:problem1} (Lyapunov weight $V=100$ and CSI feedback delay $T_{\text{delay}}=10$ms).}
    \label{fig:v2u_rate_different_v2v_number}
    \vspace{-3mm}
\end{figure}

\begin{figure}[t!]
    \centering
    \includegraphics[width=.42\textwidth]{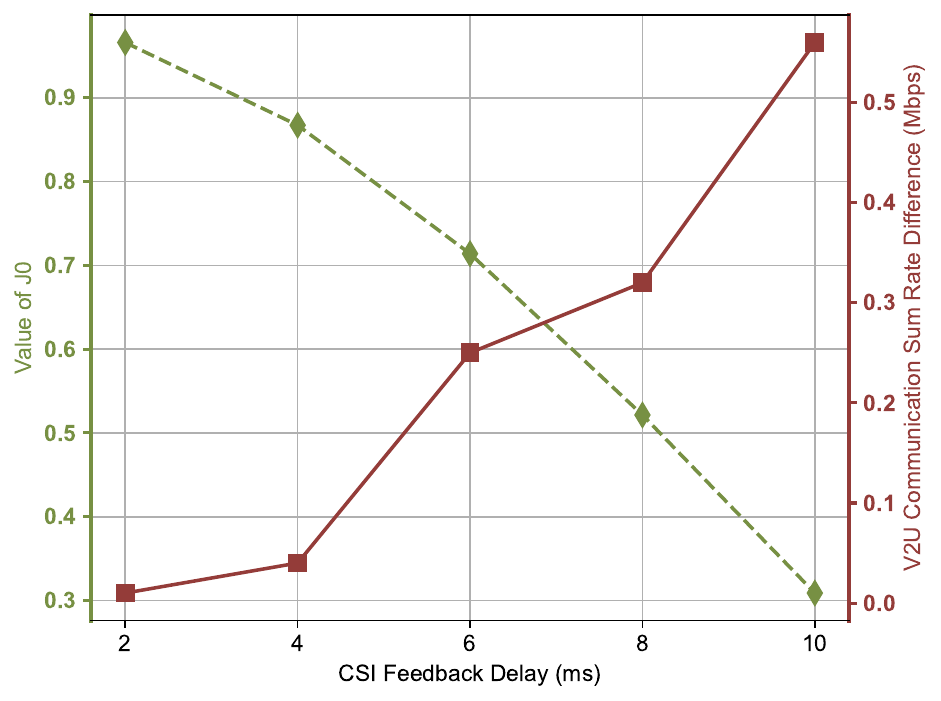}
    \vspace{-3mm}
    \caption{Impact of the value of CSI feedback delay on the V2U communication sum rate in~\eqref{eq:problem1} (the number of V2V links $K=10$ and Lyapunov weight $V=100$).}
    \label{fig:v2u_rate_different_csi_value}
\end{figure}

\subsubsection{Effect of the Number of V2V Communications}
In Fig.~\ref{fig:v2u_rate_different_v2v_number}, we illustrate the impact of incrementally increasing the number of V2V communication links on the V2U communication sum rate. The results show a clear downward trend in the V2U communication sum rate as the number of V2V communication links increases. This is because, as more V2V links are introduced into the network, a larger portion of them begin to reuse the spectrum resources originally allocated to the V2U links. This spectrum reuse introduces additional interference to V2U transmissions, thereby degrading overall V2U communication performance. Overall, the proposed D3PG outperforms the other algorithms, achieving performance improvements of 4.37\% over D3PG-WCSI, 15.34\% over DDPG, and 30.67\% over H-DDQN when the number of V2V links is set to 6. Additionally, when the number of V2V links is 10, D3PG outperforms D3PG-WCSI by 6.39\%, DDPG by 12.55\%, and H-DDQN by 23.25\%.

\subsubsection{Effect of the Value of CSI Feedback Delay}
In Fig.~\ref{fig:v2u_rate_different_csi_value}, we jointly depict the value of the Bessel function $J_0$ and the corresponding difference in V2U communication sum rate between D3PG and D3PG-WCSI under varying CSI feedback delays. The results show that as the delay increases from 2 ms to 10 ms, the value of $J_0$ decreases monotonically, indicating that the outdated CSI becomes increasingly decorrelated from the true channel state. Simultaneously, the performance gap in V2U communication between D3PG and D3PG-WCSI gradually widens. This is because, when the delay is small (e.g., 2 ms), $J_0$ remains close to 1, meaning the outdated CSI still closely approximates the actual channel state, resulting in negligible performance differences. However, as the delay grows, D3PG-WCSI suffers from poor decision-making due to delayed observations, while D3PG compensates for outdated CSI through a Gauss–Markov-based model. This analysis confirms that accounting for CSI aging is essential for robust decision-making and maintaining communication performance in high-mobility UAV-assisted vehicular networks.

% \begin{table}[htbp]
% \centering
% \footnotesize
% \caption{Comparison of algorithm running time per time slot (millisecond).}
% % \vspace{-2mm}
% \label{table2}
% \rowcolors{1}{lightblue!30}{white}
% \begin{tabular}{|c|c|c|c|c|c|}
% \hline
% \textbf{Number of Users} & \textbf{10} & \textbf{12} & \textbf{14} & \textbf{16} & \textbf{18} \\ \hline\hline
% T2DRL & 1.112 & 1.162 & 1.197 & 1.226 & 1.272\\ \hline
% DDPG-based T2DRL & 0.302 & 0.317 & 0.324 & 0.339 & 0.351 \\ \hline
% SCHRS & 624.9 & 769.7 & 844.1 & 973.2 & 1128.7 \\ \hline
% \end{tabular}
% \end{table}

\begin{figure}[t!]
    \centering
    \includegraphics[width=.4\textwidth]{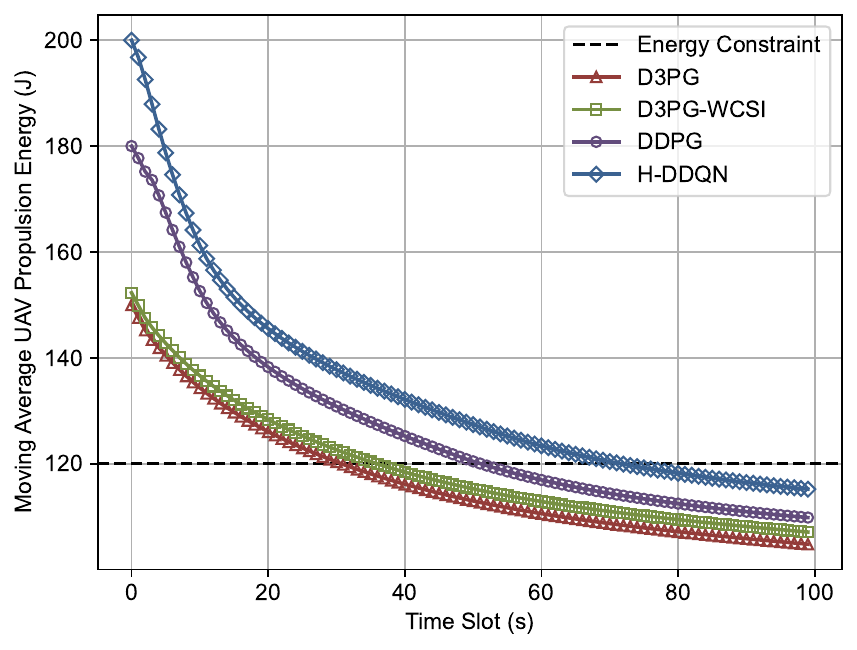}
    \vspace{-3mm}
    \caption{UAV propulsion energy consumption over time slots (the number of V2V links $K=10$, Lyapunov weight $V=100$, and CSI feedback delay $T_{\text{delay}}=10$ms).}
    \label{fig:UAV_propulsion_energy_different_slots}
    \vspace{-3mm}
\end{figure}

\begin{figure}[t!]
    \centering
    \includegraphics[width=.41\textwidth]{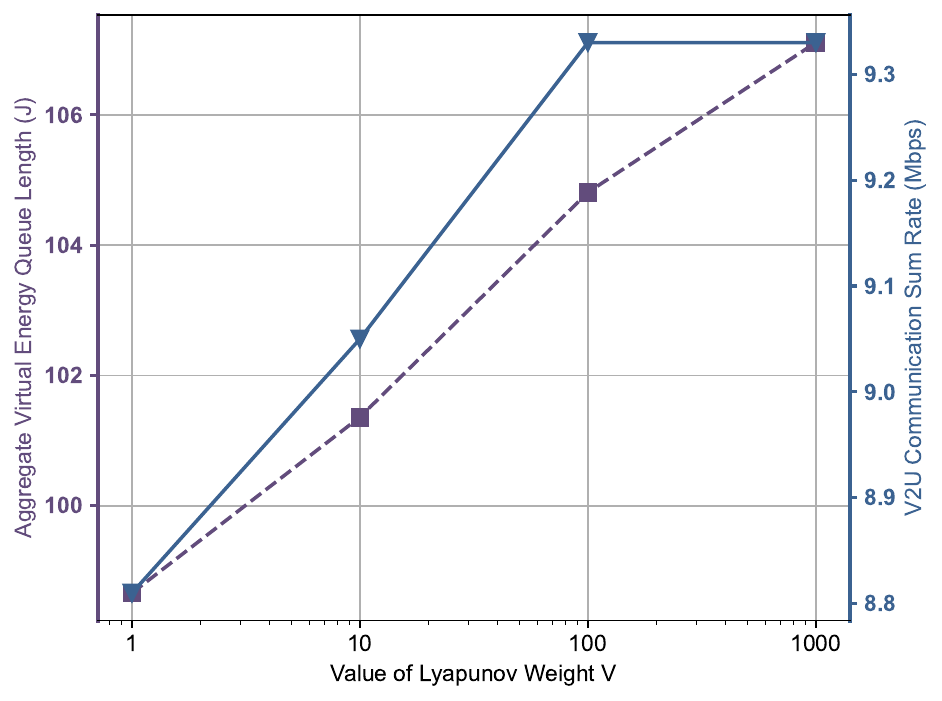}
    \vspace{-3mm}
    \caption{V2U communication sum rate in~\eqref{eq:problem1} and aggregate virtual energy queue length versus parameter $V$ (the number of V2V links $K=10$ and CSI feedback delay $T_{\text{delay}}=10$ms).}
    \label{fig:different_V}
\end{figure}

\subsubsection{UAV Energy Consumption Analysis} In Fig.~\ref{fig:UAV_propulsion_energy_different_slots}, we illustrate the moving average energy consumption of the UAV ($\frac{1}{t}\sum_{\tilde{t}=1}^{t}P(\tilde{t})\Delta$) within the considered flight duration. The results show that the long-term UAV propulsion energy consumption remains below the predefined threshold for all methods, thereby ensuring the UAV’s operational endurance. This demonstrates that, by decomposing the long-term optimization problem \eqref{eq:problem1} into per-slot subproblems \eqref{eq:problem2}, the proposed Lyapunov optimization framework not only optimizes the V2U communication sum rate but also satisfies the long-term UAV energy consumption constraint in $\mathcal{C}8$. Overall, the proposed D3PG reduces the moving average energy consumption by 2.15\%, 4.58\%, and 9.02\% compared to D3PG-WCSI,   DDPG, and H-DDQN, respectively.

\subsubsection{Effect of the Value of the Lyapunov Weight $V$} In Fig.~\ref{fig:different_V}, we present the impact of the Lyapunov control parameter $V$ on the tradeoff between the aggregate virtual energy queue length ($\frac{1}{T}\sum_{\tilde{t}=1}^{T}P(\tilde{t})\Delta$) and the V2U communication sum rate. The results show that both the aggregate virtual energy queue length and the V2U communication sum rate increase with the Lyapunov weight $V$. This is because a larger $V$ places greater emphasis on maximizing the V2U communication rate in the Lyapunov drift-plus-penalty function defined in~\eqref{eq:problem2}. However, this improvement comes at a cost: the growing length of the virtual energy queue indicates a higher likelihood of energy constraint violations. This is because the system becomes more aggressive in pursuing communication performance as $V$ increases. Notably, when $V$ becomes sufficiently large (e.g., increasing from 100 to 1000), the V2U sum rate plateaus. This saturation effect occurs because the system reaches a performance ceiling beyond which further increases in $V$ no longer yield additional V2U sum rate gains.

\begin{table}[htbp]
\centering
\footnotesize
\caption{Comparison of algorithm running time per time slot (milliseconds).}
% \vspace{-2mm}
\label{table3}
\rowcolors{1}{white}{gray!20}
\begin{tabular}{|c|c|c|c|c|c|}
\hline
\textbf{Number of V2V Links} & \textbf{6} & \textbf{7} & \textbf{8} & \textbf{9} & \textbf{10} \\ \hline\hline
D3PG & 2.64 & 2.76 & 3.18 & 3.32 & 3.34\\ \hline
DDPG & 0.45 & 0.46 & 0.51 & 0.65 & 0.66 \\ \hline
H-DDQN & 0.51 & 0.63 & 0.74 & 0.82 & 0.96 \\ \hline
\end{tabular}
\end{table}

\subsubsection{Algorithm Running Time Performance} Table~\ref{table3} presents the impact of the number of V2V links on the algorithm's running time per time slot. D3PG-WCSI is excluded from the comparison, as it only differs in whether CSI feedback delay is considered, without modifying any algorithmic modules. The results show that the running time of H-DDQN is higher than that of DDPG, mainly due to the use of the Hungarian algorithm for channel allocation, which involves matrix operations with a computational complexity of $\mathcal{O}(K^3)$. Additionally, D3PG incurs the highest running time, primarily due to the added reverse process, which generates actions through a step-wise denoising procedure. Nevertheless, given that D3PG achieves the highest V2U communication sum rate, we conclude that it offers superior performance with only a modest increase in computational complexity.

\vspace{-3mm}
\section{Conclusion and Future Works} \label{sec:conclusion}
\vspace{-.15mm}
In this paper, we have offered new insights into low-altitude economy networking by exploring intelligent UAV-assisted V2X communication strategies aligned with UAV energy efficiency. Specifically, we have addressed the problem of joint channel allocation, power control, and flight altitude adjustment in UAV-assisted vehicular networks with CSI feedback delay. We have integrated Lyapunov optimization with the proposed D3PG algorithm to ensure long-term energy efficiency while substantially enhancing V2U communication performance. We have proposed a D3PG algorithm that incorporates diffusion models into its actor network, effectively addressing the exploration–exploitation trade-off in conventional DRL while enhancing decision-making robustness in dynamic environments with CSI feedback delay through conditioning on real-time environmental features. For the future research, we will extend the proposed framework to multi-UAV scenarios, where inter-UAV coordination poses additional challenges in distributed decision-making and energy management.
\vspace{-3mm}

% Future research could further explore the potential of utilizing an expert dataset which can be obtained offline through the brute-force search method, to conduct the forward process of diffusion models. This data set includes the optimal solutions for DNN partitioning and task offloading. Subsequently, a form of supervised learning could be applied to train the diffusion model to fit the action distribution generated by the reverse process and the expert data. Furthermore, investigations of competitions and cooperations among multiple RSUs for task acquisition is an enticing direction. 
% \vspace{-3mm}

\begingroup
% \onecolumn

\appendices

\section{}\label{app:virtual_queue}
Given the virtual energy queue definition:
\vspace{-.5mm}
\begin{equation}\label{eq:a_app1}
Q(t+1)=\max \Big\{Q(t)+P(t)\Delta-E^{\text{U}}_{\text{th}},0\Big\},
\end{equation}
we derive the inequality:
\vspace{-.5mm}
\begin{equation}\label{eq:a_app2}
Q(t+1)\geq Q(t)+P(t)\Delta-E^{\text{U}}_{\text{th}}.
\end{equation}
Applying sample path analysis~\cite{10460374} and summing over $t=1,...,T$ yields:
\vspace{-.5mm}
\begin{equation}\label{eq:a_app3}
Q(T)\geq Q(1)+\sum_{t=1}^T P(t)\Delta-TE^{\text{U}}_{\text{th}}.
\end{equation}
For finite $Q(T)$ and $Q(1)$, taking $T\to \infty$ gives:
\vspace{-.5mm}
\begin{align}\label{eq:a_app4}
&\lim_{T\to \infty} \frac{1}{T}\sum_{t=1}^T P(t)\Delta \nonumber \\ &\leq\lim_{T\to \infty}\Big(\frac{Q(T)-Q(1)}{T}+ E^{\text{U}}_{\text{th}}\Big)  =E^{\text{U}}_{\text{th}}.
\end{align}

\section{}\label{app:upperbound}
Beginning with the queue dynamics in~\eqref{eq:queue_update} and applying the inequality $\Big(\max\{a+b-c,0\}\Big)^2 \leq (a+b-c)^2$, we obtain:
\vspace{-.5mm}
\begin{align}\label{eq:b_app1}
\Big(Q(t+1)\Big)^2\leq\Big(Q(t)+P(t)\Delta-E^{\text{U}}_{\text{th}}\Big)^2,
\end{align}
% namely,
% \vspace{-.5mm}
% \begin{align}\label{eq:b_app2}
% \Big(Q(t+1)\Big)^2&\leq\Big(Q(t)\Big)^2+2Q(t)\Big(P(t)\Delta-E^{\text{U}}_{\text{th}}\Big)\nonumber\\
% &+\Big(P(t)\Delta-E^{\text{U}}_{\text{th}}\Big)^2,
% \end{align}
Expanding this relationship yields:
\vspace{-.5mm}
\begin{align}\label{eq:b_app2}
\frac{\Big(Q(t+1)\Big)^2-\Big(Q(t)\Big)^2}{2}&\leq Q(t)\Big(P(t)\Delta-E^{\text{U}}_{\text{th}}\Big)\nonumber\\
&+\frac{1}{2}\Big(P(t)\Delta-E^{\text{U}}_{\text{th}}\Big)^2.
\end{align}
This leads to the Lyapunov drift bound:
\vspace{-.5mm}
\begin{align}\label{eq:b_app3}
\Delta L\Big(Q(t)\Big)&\leq \mathbb{E}\Big\{ Q(t)\Big(P(t)\Delta-E^{\text{U}}_{\text{th}}\Big)\mid Q(t)\Big\} \nonumber\\
&+\frac{1}{2}\Big(P(t)\Delta-E^{\text{U}}_{\text{th}}\Big)^2.
\end{align}
Consequently, we derive the complete Lyapunov drift-plus-penalty expression:
\vspace{-.5mm}
\begin{align}\label{eq:b_app4}
&D\Big(Q(t)\Big)\leq \mathbb{E}\Big\{ Q(t)\Big(P(t)\Delta-E^{\text{U}}_{\text{th}}\Big)\mid Q(t)\Big\}\nonumber\\
&-V\mathbb{E}\Big\{\frac{1}{M} \sum_{m\in\mathcal{M}} R^{\text{U}}_{m}(t)\mid Q(t)\Big\}
+\frac{1}{2}\Big(P(t)\Delta-E^{\text{U}}_{\text{th}}\Big)^2.
\end{align}

\bibliographystyle{IEEEtran}

\vspace{-14mm}

\end{document}